\documentclass[twocolumn, pra, superscriptaddress]{revtex4-1}
\usepackage{amssymb}
\usepackage{amsmath}
\usepackage{epsfig}
\usepackage{graphics, graphicx}
\usepackage{bbold}
\usepackage{psfrag}
\usepackage{mathcomp}
\usepackage{mathrsfs}
\usepackage{subfigure}
\usepackage{verbatim}
\usepackage{color}
\usepackage[colorlinks,citecolor=blue]{hyperref}
\def\cp#1{\mathbf{#1}}

\begin{document}
	\title{Quantum fluctuations in a quartet superfluid of two-dimensional Fermi mixtures}
	
	\author{Wei Wang}
	\affiliation{Beijing National Laboratory for Condensed Matter Physics, Institute of Physics, Chinese Academy of Sciences, Beijing, 100190, China}
	\affiliation{School of Physical Sciences, University of Chinese Academy of Sciences, Beijing 100049, China}
		\author{Yupeng Wang}
	\affiliation{Beijing National Laboratory for Condensed Matter Physics, Institute of Physics, Chinese Academy of Sciences, Beijing, 100190, China}
	
	\author{Xiaoling Cui}
	\affiliation{Beijing National Laboratory for Condensed Matter Physics, Institute of Physics, Chinese Academy of Sciences, Beijing, 100190, China}
	
	\begin{abstract}
We study quantum fluctuations in a quartet superfluid (QSF) of two-dimensional (2D) fermion mixtures with mass imbalance. Here QSF is a high-order superfluid that corresponds to the condensation of ($1+3$) clusters, each consisting of a light fermion and three heavy ones. By incorporating the Gaussian fluctuations respecting dominant four-body correlations in this system, our theory successfully recovers the leading universal logarithmic contribution to the 2D equation of state in the deep binding regime, thereby offering a correct physical picture of quartet clusters behaving as composite bosons. By extending the Gaussian fluctuation theory from pairing to quartet superfluids, our results shed light on quantum fluctuations in general fermion superfluids with arbitrarily high-order correlations.
	\end{abstract}
	
	\date{\today}
	\maketitle
	
\section{Introduction}
Fermion superfluidity in the strong coupling regime is one of the most attractive while challenging subjects in modern physics,  with the main challenge arising from the absence of a small parameter for perturbative treatments. In the literature, strong coupling theory of fermion superfluids has generated great interest across various fields, including condensed matter, nuclear, and atomic and molecular physics\cite{review_1,review_2,review_nuclear1,review_nuclear2}. In ultracold atoms, pairing superfluids in both three-dimensional (3D)\cite{3D1,3D2,3D3,3D4,3D5,3D6,3D7,3D8,3D9,3D10,3D11,3D12,3D13,3D14,3D15,3D16} and two-dimensional (2D)\cite{2D1,2D2,2D3,2D4,2D5,2D6,2D7,2D8,2D9,2D10,2D11} fermion systems have been successfully realized, providing an ideal platform for testing various strong coupling theories. 
For these pairing superfluids, although the mean-field theory is able to provide a qualitative description during the Bardeen-Cooper-Schrieffer (BCS) to Bose-Einstein condensation (BEC) crossover, it fails to provide quantitative agreement with experimental observations in the strong coupling regime. In this regime, quantum fluctuations become crucially important, as incorporated in quantum Monte Carlo simulations\cite{MC3D1,MC3D2,MC3D3,MC3DT1,MC3DT2,MC3DT3,MC2D1,MC2D2,MC2D3,MC2DT1,MC2DT2,MC2DT3,MCmassimba1,MCmassimba2,MCmassimba3} and various many-body T-matrix approaches\cite{NSR,Melo1,Haussmann1,Melo2,Ohashi1,Strinati1,ChenQijin,HuHui1,Ohashi2,Haussmann2,Randeria,QF2D0,Ohashi3,QF2DBKT,Ohashi4,Strinati2}. In particular, the fluctuation effect can be seen transparently in the deep BEC regime where paired fermions can be viewed as composite bosons, namely, quantum fluctuations must be included to correctly describe their effective interactions, equation of state (EOS), and critical temperature for both 3D superfluid transition\cite{Melo1,Haussmann1} and 2D Berezinskii–Kosterlitz–Thouless (BKT) transition\cite{QF2DBKT,Salasnich}. Especially in two dimensions, quantum fluctuations become extremely important in order to recover the logarithmic dependence of EOS\cite{QF2D0}, a behavior that cannot be produced at the mean-field level.

Despite a comprehensive understanding of quantum fluctuations in pairing superfluids, their effects in a high-order fermion superfluid beyond the pairing framework remain largely unknown. The leading example of high-order superfluid is the quartet superfluid (QSF) that corresponds to the condensation of four-fermion clusters and has been revealed in various physical systems in the literature\cite{Wu1,quartet_nuclear1,quartet_nuclear2,quartet_nuclear3,quartet_nuclear4,Tajima1,Tajima2,Babaev1,Kivelson,Babaev2,Yao1,Fu,Yao2,Babaev3,charge4e_expt,HuJiangping,WangYuxuan,Wu2,Babaev4,Jian}. Among these theoretical studies, most are based on mean-field-type treatments or phenomenological Ginzburg-Landau or London models with various fluctuations, while works that simultaneously keep the full internal momentum structure and incorporate quantum fluctuations are rather limited. 

Recently, QSF has been shown to emerge in a simple setup of 2D Fermi-Fermi mixtures with mass imbalance\cite{Cui2}, where a light fermion can bind with three heavy ones to form a ($1+3$) cluster bound state as a unit quartet in vacuum\cite{Blume, Petrov, Parish, Cui1}. Based on a variational ansatz reflecting the mean-field condensation of quartet clusters, QSF has been identified as the ground state in a wide parameter region of strong coupling strength and large mass imbalance\cite{Cui2}. Nevertheless, since this ansatz does not include any quantum fluctuations, it can only provide a qualitative description of QSF, analogous to the mean-field treatment of pairing superfluids. To retrieve a correct description of quartet clusters in the deep binding limit, one must carefully incorporate fluctuation effects in this high-order superfluid, especially for two dimensions where quantum fluctuations are much enhanced compared to three dimensions. However, this problem is highly nontrivial since the quantum fluctuations have to respect the dominant high-order correlations in the corresponding mean-field background.

In this work, we attempt to evaluate quantum fluctuations in QSF, which is dominated by ($1+3$) correlation well beyond the conventional ($1+1$) correlation in pairing superfluids. Motivated by the success of Gaussian fluctuation theory in pairing superfluids\cite{Melo2,HuHui1,Randeria,QF2D0,QF2DBKT,Salasnich}, here we establish an extended Gaussian framework for QSF by incorporating the dominant four-body correlations using the functional path integral approach. 
At the mean-field level, our theory directly reproduces the gap equation obtained from the variational method at zero temperature\cite{Cui2}, and can also be extended to treat QSF at finite temperatures. When incorporating quantum fluctuations, our theory can capture the leading logarithmic term of the 2D EOS in the deep binding limit, thereby offering a correct physical picture of quartet clusters behaving as composite bosons in this regime. By extending the Gaussian fluctuation theory from pairing superfluids to QSF, our results shed light on quantum fluctuations in general fermion superfluids with arbitrarily high-order correlations.

The remainder of this paper is organized as follows. Section~\ref{model} is devoted to the basic model of our system and the main formalism based on the functional path integral approach. Sections~\ref{mf} and \ref{QF} are respectively devoted to the mean-field analysis and Gaussian fluctuation theory of QSF. Finally we summarize our work in Sec.~\ref{summary}.

\section{Model and formalism} \label{model}
	
We write down the Hamiltonian of mass-imbalanced Fermi mixtures in 2D ($\hbar=1$):
\begin{equation}
	{\cal H}=\sum_{{\cp k}} \left(\epsilon^l_{{\cp k}} l_{{\cp k}}^{\dagger} l_{{\cp k}} + \epsilon^h_{{\cp k}} h_{{\cp k}}^{\dagger} h_{{\cp k}}\right) +\frac{g}{S} \sum_{\mathbf{q}, {\cp k}, {\cp k}^{\prime}} l_{\mathbf{q}-{\cp k}}^{\dagger} h_{{\cp k}}^{\dagger} h_{{\cp k}^{\prime}} l_{\mathbf{q}-{\cp k}^{\prime}}. \label{eq:H}
\end{equation}
Here $h_{{\cp k}}^{\dagger}$ ($l_{{\cp k}}^{\dagger}$) is the creation operator of a heavy (light) fermion at momentum ${\cp k}$ with mass $m_h$ ($m_l$); the single-particle energy is $\epsilon^{h,l}_{{\cp k}}={\cp k}^2/(2m_{h,l})$;
the bare coupling $g$ is renormalized via $1/g=-1/S \sum_{{\cp k}}1/(\epsilon^l_{{\cp k}}+\epsilon^h_{{\cp k}}+E_{2b})$, where $S$ is the system area and $E_{2b}=(2m_r a^2)^{-1}$ is the two-body binding energy with reduced mass $m_r=m_lm_h/(m_l+m_h)$ and 2D scattering length $a$. Same as Ref.\cite{Cui2}, in this work we consider the heavy-light number ratio as $N_h:N_l=3:1$, which is the most favorable situation for QSF. Moreover, we define a momentum unit $k_F=\sqrt{4\pi N_Q/S}$, with $N_Q=N_l=N_h/3$ the number of quartets.
	
In Ref.\cite{Cui2}, we have proposed a variational ansatz for QSF at zero temperature ($T=0$) based on the idea of quartet condensation. Here we reproduce the results in Ref.\cite{Cui2} using the functional path integral approach. Considering that each light fermion binds with three heavy ones to form a quartet unit, we introduce 
a composite fermion operator $H^{\dag}_{{\cp k}_1{\cp k}_2{\cp k}_3}\equiv h^{\dag}_{{\cp k}_1}h^{\dag}_{{\cp k}_2} h^{\dag}_{{\cp k}_3}$, and then write the grand canonical Hamiltonian as 
\begin{align}
    {\cal H}&=\sum_{{\cp k}} \xi^l_{{\cp k}} l_{{\cp k}}^{\dagger} l_{{\cp k}} + \sum_{\{{\cp k}_1{\cp k}_2{\cp k}_3\}} \xi^H_{{\cp k}_1{\cp k}_2{\cp k}_3} H_{{\cp k}_1{\cp k}_2{\cp k}_3}^{\dagger} H_{{\cp k}_1{\cp k}_2{\cp k}_3} \nonumber\\
&+\frac{g}{S} \sum_{\mathbf{q}, {\cp k}, {\cp k}^{\prime}} \sum_{\{{\cp k}_2{\cp k}_3\}} l_{\mathbf{q}-{\cp k}-{\cp k}_2-{\cp k}_3}^{\dagger} H_{{\cp k}{\cp k}_2{\cp k}_3}^{\dagger} H_{{\cp k}^{\prime}{\cp k}_2{\cp k}_3} l_{\mathbf{q}-{\cp k}^{\prime}-{\cp k}_2-{\cp k}_3}, \nonumber\\
\label{eq:H2}
\end{align}
where $\xi^l_{{\cp k}}=\epsilon^l_{{\cp k}}-\mu_l$ and $\xi^H_{{\cp k}_1{\cp k}_2{\cp k}_3}=\epsilon^h_{{\cp k}_1}+\epsilon^h_{{\cp k}_2}+\epsilon^h_{{\cp k}_3}-\mu_{H}$, with $\mu_{l/H}$ the corresponding chemical potentials; the notation $\left\{\mathbf{k}_1\mathbf{k}_2\mathbf{k}_3\right\}$ ($\left\{\mathbf{k}_2\mathbf{k}_3\right\}$) denotes a summation over three (two) momentum indices with the restrictions that they are mutually distinct and that the six (two) permutations of the same set are counted only once. 

Strictly speaking, the composite fermion operators do not satisfy canonical fermionic anticommutation relation. For instance, we have
\begin{eqnarray}
\left\{H^{\dag}_{\mathbf{k}_1\mathbf{k}_2\mathbf{k}_3},H_{\mathbf{k}_3\mathbf{k}_2\mathbf{k}_1}\right\}&=&1-n_{\mathbf{k}_1}-n_{\mathbf{k}_2}-n_{\mathbf{k}_3}\nonumber\\
&+&n_{\mathbf{k}_1}n_{\mathbf{k}_2}+n_{\mathbf{k}_1}n_{\mathbf{k}_3}+n_{\mathbf{k}_2}n_{\mathbf{k}_3}, \label{permutation}
\end{eqnarray}
with $n_{\mathbf{k}}=h^{\dag}_{{\cp k}}h_{{\cp k}}$ the number operator of heavy fermion at momentum ${\cp k}$. In the deep binding limit, $n_{\mathbf{k}}\to0$, so the number correction in (\ref{permutation}) is negligible and the anticommutation relation is recovered. Physically, such a correction is due to the Pauli-blocking effect of heavy fermions, or equivalently the momentum overlap between two composite fermion operators. Similarly, we have $\left\{H^{\dag}_{\mathbf{k}_1\mathbf{k}_2\mathbf{k}_3},H_{\mathbf{k}'_3\mathbf{k}'_2\mathbf{k}'_1}\right\}\neq 0$ if $\{\mathbf{k}_1\mathbf{k}_2\mathbf{k}_3\}$  and $\{\mathbf{k}'_1\mathbf{k}'_2\mathbf{k}'_3\}$ have one or two momenta in common. In view of this, our theory is expected to work well only in the deep binding limit where the Pauli-blocking effect can be safely neglected. This effect has also been discussed previously under the variational framework\cite{Cui2}.  

Introducing an auxiliary pairing field $\phi_{{\cp k}_2{\cp k}_3}({\cp q},\tau)$ and applying the Hubbard-Stratonovich transformation, we get the action in Gaussian form:
\begin{widetext}
\begin{eqnarray}
S\left[\bar{\phi},\phi,\bar{\Psi},\Psi \right]=-\int_{0}^{\beta}d\tau\left\{ \sum_{\left\{\mathbf{k}\right\}}\sum_{\left\{\mathbf{k'}\right\}}\bar{\Psi}(\left\{\mathbf{k}\right\},\tau)G^{-1}(\left\{\mathbf{k}\right\},\left\{\mathbf{k'}\right\},\tau) \Psi (\left\{\mathbf{k'}\right\},\tau)+\frac{S}{g}\sum_{\mathbf{q}}\sum_{\{\mathbf{k_2k_3}\}}\bar{\phi}_{\{\mathbf{k_2k_3}\}}(\mathbf{q},\tau )\phi_{\mathbf{k_2k_3}}(\mathbf{q},\tau)\right\},  \label{S}
\end{eqnarray}
where $\{ \mathbf{k_1k_2k_3}\}$ is simplified as $\{ \mathbf{k}\}$;  the Nambu-Gor’kov spinor reads $\Psi(\left\{\mathbf{k}\right\},\tau)=[H_{\mathbf{k_1k_2k_3}}(\tau),\bar{l}_{\mathbf{-k_1-k_2-k_3}}(\tau)]^{T}$; $\tau$ is the imaginary time ranging from $0$ to $\beta=1/(k_BT)$. The inverse Green's function is given by
\begin{eqnarray}
G^{-1}(\left\{\mathbf{k}\right\},\left\{\mathbf{k'}\right\},\tau)= \begin{pmatrix} -(\partial _{\tau}+\xi_{\mathbf{k_1k_2k_3} }^h)\delta_{\{\cp k\}\{\cp k'\} }  & \Phi (\left \{ \mathbf{k}  \right \},\left \{ \mathbf{k'}  \right \},\tau ) \\ \bar{\Phi} (\left \{ \mathbf{k'}  \right \},\left \{ \mathbf{k}  \right \},\tau ) &-(\partial _{\tau}-\xi_{\mathbf{-k_1-k_2-k_3}}^l)\delta_{\{\cp k\}\{\cp k'\} } \end{pmatrix},
\end{eqnarray}
with 
\begin{equation}
\Phi (\left \{ \mathbf{k}  \right \},\left \{ \mathbf{k'}  \right \},\tau)=\phi_{\mathbf{k_2k_3}}(\mathbf{q},\tau)-\phi_{\mathbf{k_1k_3}}(\mathbf{q},\tau)+\phi_{\mathbf{k_1k_2}}(\mathbf{q},\tau).
\end{equation}
\end{widetext}
Here the bosonic field $\Phi (\left \{ \mathbf{k}  \right \},\left \{ \mathbf{k'}  \right \},\tau)$ corresponds to quartet annihilation with total momentum $\mathbf{q}=\mathbf{k}_1+\mathbf{k}_2+\mathbf{k}_3-\mathbf{k}_1'-\mathbf{k}_2'-\mathbf{k}_3'$. Note that the two-body interactions require that the two momentum triples in $\Phi$, i.e., $\{\mathbf{k}\}$ and $\{\mathbf{k'}\}$, share at least two momenta in common. For instance, we can have $\mathbf{k}_1=\mathbf{k}'_1$, $\mathbf{k}_2=\mathbf{k}'_2$, and $\mathbf{q}=\mathbf{k}_3-\mathbf{k}'_3$.

After integrating out the fermion fields, we obtain the effective action:
\begin{equation}
S_{\mathrm{eff}}=-\ln\det G^{-1}-\frac{S}{g}\sum_{q}\sum_{\{\mathbf{k_2k_3}\}}\bar{\phi}_{\mathbf{k_2k_3}}(q)\phi_{\mathbf{k_2k_3}}(q),\label{effS}
\end{equation}
where $q=(\mathbf{q},\mathrm{i}v_n)$ and $v_n=2n\pi/\beta$ is the bosonic Matsubara frequency. The partition function then reads $Z=\int \mathscr{D}\left [ \bar{\phi },\phi \right]\exp \left[- S_{\mathrm{eff}} \right]$.

\section{Mean-field analysis} \label{mf}

For the quartet condensation at ${\cp q}=0$, one has $\{\mathbf{k}\}=\{\mathbf{k'}\}$ in Eq.(\ref{S}), which gives rise to the mean-field action 
\begin{equation}
S_0=-\ln\det G_0^{-1}-\frac{S}{g}\sum_{\{\mathbf{k_2k_3}\}}\Delta_{\mathbf{k_2k_3}}^2,
\end{equation}
where we denote $\phi_{\mathbf{k_2k_3}}({\cp q}=0,\tau)\equiv \Delta_{\mathbf{k_2k_3}}$, and 
\begin{equation}
G_0^{-1} = \begin{pmatrix} \mathrm{i}\omega_n-\xi_{\mathbf{k_1k_2k_3} }^H & \Delta_{\mathbf{k_1k_2k_3}} \\ \Delta_{\mathbf{k_1k_2k_3}} &\mathrm{i}\omega_n+\xi_{\mathbf{-k_1-k_2-k_3}}^l \end{pmatrix},  
\end{equation}   
with a combined pairing amplitude $\Delta_{\mathbf{k_1k_2k_3}}\equiv\Delta_{\mathbf{k_2k_3}}-\Delta_{\mathbf{k_1k_3}}+\Delta_{\mathbf{k_1k_2}}$ and fermionic Matsubara frequency $\omega_n=(2n+1)\pi/\beta$.
Then the thermodynamic potential $\Omega_0=S_0/\beta$ can be simplified as
\begin{align}
    \Omega_0 &=\sum_{\left\{\mathbf{k}\right\}}(\xi_{\mathbf{k_1k_2k_3}}^+-\sqrt{\xi_{\mathbf{k_1k_2k_3}}^{+2}+\Delta_{\mathbf{k_1k_2k_3}}^2})\nonumber\\&-\frac{1}{\beta}[\sum_{\left\{\mathbf{k}\right\}}\ln(1+e^{\beta E_{\mathbf{k_1k_2k_3}}^-}) +\sum_{\left\{\mathbf{k}\right\}}\ln(1+e^{-\beta E_{\mathbf{k_1k_2k_3}}^+})]\nonumber\\&- \frac{S}{g}\sum_{\{\mathbf{k_2k_3}\}}\Delta_{\mathbf{k_2k_3}}^2,   \label{Omega}
\end{align}
where $\xi_{\mathbf{k_1k_2k_3}}^{\pm}=(\xi_{\mathbf{-k_1-k_2-k_3}}^l\pm \xi_{\mathbf{k_1k_2k_3}}^H)/2$ and $E_{\mathbf{k_1k_2k_3}}^\pm=-\xi_{\mathbf{k_1k_2k_3}}^{-} \pm \sqrt{\xi_{\mathbf{k_1k_2k_3}}^{+2}+\Delta_{\mathbf{k_1k_2k_3}}^2 }$.
The saddle point equation $\partial \Omega_0/\partial \Delta_{\mathbf{k_2k_3}}=0$ then leads to 
\begin{align}
\frac{S}{g}\Delta_{\mathbf{k_2k_3}}&=\sum_{\mathbf{k_1}} [n_F(E_{\mathbf{k_1k_2k_3}}^+)-n_F(E_{\mathbf{k_1k_2k_3}}^-)]\nonumber\\&\times\frac{\Delta_{\mathbf{k_1k_2k_3}}}{2\sqrt{\xi_{\mathbf{k_1k_2k_3}}^{+2}+\Delta_{\mathbf{k_1k_2k_3}}^2}}.
\label{gapequation}
\end{align} 
The occupation number is $n_F(E)=1/(e^{\beta E}+1)$.
The number equations from $N_{l}=-\partial \Omega_0/\partial \mu_{l}$ and $N_{h}=-3\partial \Omega_0/\partial \mu_{H}$ read
\begin{align}
    N_l&=\frac{1}{2}\sum_{\{\mathbf{k}\}}[n_F(-E_{\mathbf{k_1k_2k_3}}^-)(1+\frac{\xi_{\mathbf{k_1k_2k_3}}^+}{\sqrt{\xi_{\mathbf{k_1k_2k_3}}^{+2}+\Delta_{\mathbf{k_1k_2k_3}}^2}})
    \nonumber\\&+n_F(-E_{\mathbf{k_1k_2k_3}}^+)(1-\frac{\xi_{\mathbf{k_1k_2k_3}}^+}{\sqrt{\xi_{\mathbf{k_1k_2k_3}}^{+2}+\Delta_{\mathbf{k_1k_2k_3}}^2}})], \\ N_h&=\frac{3}{2} \sum_{\{\mathbf{k}\}}[n_F(E_{\mathbf{k_1k_2k_3}}^+)(1+\frac{\xi_{\mathbf{k_1k_2k_3}}^+}{\sqrt{\xi_{\mathbf{k_1k_2k_3}}^{+2}+\Delta_{\mathbf{k_1k_2k_3}}^2}})\nonumber\\&+n_F(E_{\mathbf{k_1k_2k_3}}^-)(1-\frac{\xi_{\mathbf{k_1k_2k_3}}^+}{\sqrt{\xi_{\mathbf{k_1k_2k_3}}^{+2}+\Delta_{\mathbf{k_1k_2k_3}}^2}})].\label{N_eq}
\end{align}
The requirement $N_h=3N_l$ leads to 
\begin{equation}
\sum_{\left\{\mathbf{k}\right\}}\left[ n_F(E_{\mathbf{k_1k_2k_3}}^+)+n_F(E_{\mathbf{k_1k_2k_3}}^-)-1 \right ]=0.
\end{equation}
In principle, the above gap and number equations apply at any finite $T$.
At $T=0$ and assuming $\mu_{l/H}$ are both negative, Eqs.(\ref{gapequation}) and (\ref{N_eq}) reproduce the gap and number equations obtained from the variational approach\cite{Cui2}, where one can define the quartet chemical potential as $\mu_Q\equiv \mu_{l}+\mu_{H}<0$. Explicitly, at $T=0$ we have the gap and number equations 
 \begin{equation}
\frac{S}{g} \Delta_{\mathbf{k_2k_3}}+\sum_{\mathbf{k_1}}\frac{\Delta_{\mathbf{k_1k_2k_3}}}{2\sqrt{\xi_{\mathbf{k_1k_2k_3}}^{+2}+\Delta_{\mathbf{k_1k_2k_3}}^2}}=0,\label{gapequation_T0}
\end{equation}
\begin{equation}
N_Q=\frac{1}{2}\sum_{\mathbf{\left\{k\right\}}}(1-\frac{\xi_{\mathbf{k_1k_2k_3}}^+}{\sqrt{\xi_{\mathbf{k_1k_2k_3}}^{+2}+\Delta_{\mathbf{k_1k_2k_3}}^2}}),\label{N_eq_T0}
\end{equation}
and the thermodynamic potential 
\begin{align}
    \Omega_0 &=\sum_{\left\{\mathbf{k}\right\}}(\xi_{\mathbf{k_1k_2k_3}}^+-\sqrt{\xi_{\mathbf{k_1k_2k_3}}^{+2}+\Delta_{\mathbf{k_1k_2k_3}}^2})\nonumber\\&- \frac{S}{g}\sum_{\{\mathbf{k_2k_3}\}}\Delta_{\mathbf{k_2k_3}}^2.   \label{Omega_T0}
\end{align}

It is worth pointing out that in QSF the pairing amplitude $\Delta_{\mathbf{k_2k_3}}$ relies on two momentum indices,  instead of a constant as in the pairing superfluid. As analyzed in Ref.\cite{Cui2}, this is because after contracting a heavy-light pair in a quartet unit, there are still two momenta left to describe the motions of the remaining two heavy fermions. 
Moreover, the gap equation (\ref{gapequation_T0}) shows that $\Delta_{\mathbf{k_2k_3}}$ can couple with $\Delta_{\mathbf{k_2k}}$ or $\Delta_{\mathbf{k_3k}}$ with $\mathbf{k}$ free, but not with $\Delta_{\mathbf{kk'}}$ where both momentum labels are changed. This property can be attributed to the presence of two-body interaction: at each time of two-body scattering only one heavy fermion can change its momentum. In fact, these properties of $\Delta_{\mathbf{kk'}}$, including its double-momentum labeling and off-diagonal coupling, directly manifest the dominant four-body correlations in this system. In the limit of vanishing fermion density or strong attraction, we have $|\mu_Q|\gg |\Delta_{\mathbf{kk'}}|$, and the gap equation (\ref{gapequation_T0}) successfully recovers the Skorniakov–Ter-Martirosian (STM) equation of the four-body problem. Because both the variational approach in Ref.\cite{Cui2} and the functional integral approach in the present work have fully captured the essential four-body correlation in this system, they arrive at consistent results of Eqs.(\ref{gapequation_T0})-(\ref{Omega_T0}) at the mean-field level of QSF.

\section{Gaussian fluctuations} \label{QF}

To simplify the discussion of quantum fluctuations in QSF, we consider the ground state at $T=0$. 
Similar to the treatment of Gaussian fluctuations in pairing superfluids\cite{Melo2,HuHui1,Randeria,QF2D0,QF2DBKT,Salasnich}, here we assume the gap equation (\ref{gapequation_T0}) and the solution of $\Delta_{\mathbf{k_2k_3}}$ are kept fixed, and expand  
\begin{equation}
\phi_{\mathbf{k_2k_3}}(q)\equiv \Delta_{\mathbf{k_2k_3}}+\eta_{\mathbf{k_2k_3}}(q),\label{GFdefine}
\end{equation}
where $\eta$ is the fluctuation field. Then the action in Eq.(\ref{S}) can be expanded as $S_{\rm eff}=S_0+S_{2}$, with $S_2$ arising from quantum fluctuations:
\begin{equation}
S_2=\frac{1}{2} \sum_q \left(\begin{array}{cc}\bar{\eta}(q) & \eta(-q)\end{array}\right) \left(\begin{array}{cc}M_{11}(q) & M_{12}(q) \\M_{21}(q) & M_{22}(q)\end{array}\right) \left(\begin{array}{c}\eta(q) \\ \bar{\eta}(-q)\end{array}\right).
\end{equation}
Here the fluctuation vector is denoted as $\eta(q)=(...,\eta_{\mathbf{k_ik_j}},...)^T$, and therefore each $M_{ij}(q)$ is a large matrix defined on the $\{\mathbf{k}_i\mathbf{k}_j\}$ space. This is a dramatic difference from the pairing fluctuation case, where $\eta$ and $M_{ij}$ are single numbers describing the paired molecules\cite{Melo2,HuHui1,Randeria,QF2D0,QF2DBKT,Salasnich}.

\begin{figure}
    \centering
    \includegraphics[width=1.05\linewidth]{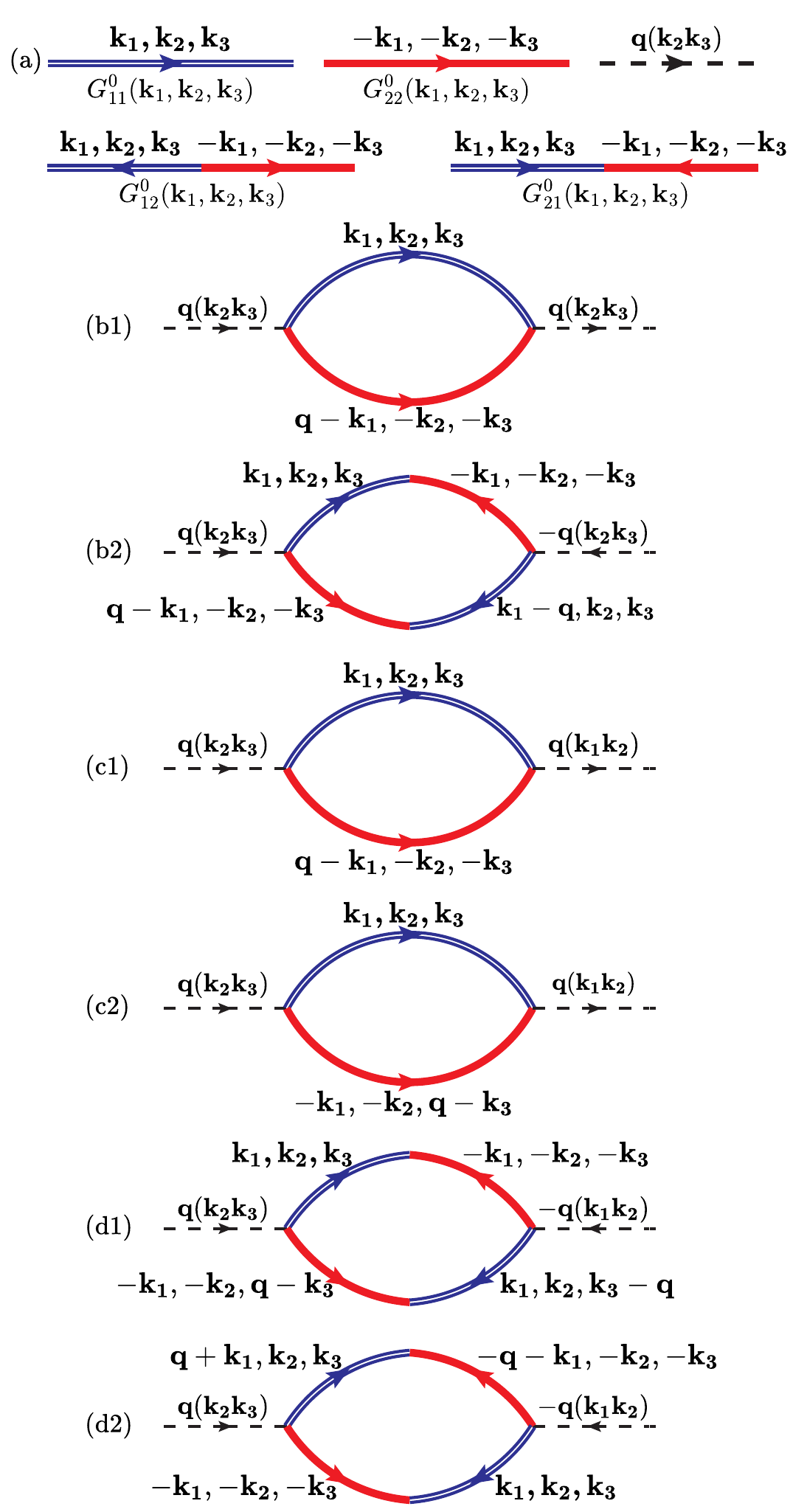}
    \caption{Diagrammatic representation of the Green's functions (a) and typical elements in the fluctuation matrix $M$ (b1-d2). In (a), we use a double blue (bold red) line with an arrow to denote the mean-field Green's function $G_{11}^0$ ($G_{22}^0$) for the heavy (light) fermion, and the lines with two arrows represent the mean-field anomalous Green's functions $G_{12}^0$ and $G_{21}^0$. The dashed line represents a quartet fluctuation field with momentum $\mathbf{q}$, within which the two unpaired heavy fermions carry momenta $\mathbf{k_2}$ and $\mathbf{k_3}$ (as labeled in the figure). (b1) and (b2) correspond to the elements in $M_{11}$ and $M_{21}$ when the two momenta $\{\mathbf{k_2k_3}\}$ are kept unchanged. These are also the typical diagrams for quantum  fluctuations in a pairing superfluid. (c1,c2) [(d1,d2)] denote the diagrams with equal contributions for $M_{11}$ [$M_{21}$] when one of the momenta is changed. These are the unique diagrams for quantum fluctuations in a quartet superfluid.}
    \label{fig_diagram}
\end{figure}

Mathematically, all elements in $M_{ij}$ can be expressed as the products of the Green's functions $G_{ij}^0$ by expanding Eq.(\ref{effS}) using Eq.(\ref{GFdefine}). The detailed expansion  is presented in Appendix \ref{sec:appendix_math_M}. For convenience, we introduce a diagrammatic representation of these fluctuation terms. After defining the diagrams for the Green's functions in Fig.\ref{fig_diagram}(a), we show the diagrams for typical elements in $M_{ij}$ in Fig.\ref{fig_diagram}(b-d). For instance, Fig.\ref{fig_diagram}(b1) shows the propagation of a quartet with center-of-mass momentum $q$ and two additional momenta $\{\mathbf{k_2k_3}\}$, denoted by $q(\mathbf{k_2k_3})$. This contributes to the diagonal term in $M_{11}$ as
\begin{align}
\left[M_{11}(q) \right]_{\mathbf{k_2k_3};\mathbf{k_2k_3}}&=-\frac{S}{g}+\frac{1}{\beta}\sum_{k_1}{'}G_{11}^0(\mathbf{k_1},\mathbf{k_2},\mathbf{k_3},\mathrm{i}\omega_n) \nonumber\\& \times G_{22}^0(\mathbf{k_1-q},\mathbf{k_2},\mathbf{k_3},\mathrm{i}\omega_n-\mathrm{i}v_n),\label{M11diag}
\end{align}
 where we denote $\sum_{k_1}^{'}\equiv\sum_{\mathbf{k_1\ne \mathbf{k_2,k_3}}}\sum_{\omega_n}$; i.e., the momentum $\mathbf{k}_1$ is summed over all values excluding $\mathbf{k}_2$ and $\mathbf{k}_3$ and the Matsubara frequency sum is also included. Fig.\ref{fig_diagram}(b2) represents the annihilation of two quartets with opposite center-of-mass momenta $q$ and $-q$, denoted by $q(\mathbf{k_2k_3})$ and $-q(\mathbf{k_2k_3})$, and gives the element
 \begin{align}
\left[M_{21}(q) \right]_{\mathbf{k_2k_3};\mathbf{k_2k_3}}&=\frac{1}{\beta}\sum_{k_1}{'}G_{21}^0(\mathbf{k_1},\mathbf{k_2},\mathbf{k_3},\mathrm{i}\omega_n)\nonumber\\ &\times G_{21}^0(\mathbf{k_1-q},\mathbf{k_2},\mathbf{k_3},\mathrm{i}\omega_n-\mathrm{i}v_n).\label{M21diag}
\end{align}
Since the two momenta  $\{\mathbf{k_2k_3}\}$ are kept fixed in both diagrams, these contributions are analogous to those in pairing superfluids that represent $M_{11}(q)$ and $M_{21}(q)$\cite{Melo2,HuHui1,Randeria,QF2D0,QF2DBKT,Salasnich}. 

In addition to these conventional diagrams, more exotic ones can emerge in QSF due to the change of momenta $\{\mathbf{k_2k_3}\}$. However, not all of them are physical because some terms correspond to changing both momenta simultaneously, while others involve a mismatch between the Green's function products and the fluctuation fields. More detailed discussions of these unphysical terms can be found in Appendix \ref{sec:appendix_math_M}. Considering the character of two-body interaction, we only keep the terms where at least one momentum of $\{\mathbf{k_2k_3}\}$ is kept unchanged in the elements of $M_{ij}$, which is therefore a sparse matrix. For instance, Fig.\ref{fig_diagram}(c1,c2) depict the annihilation of a quartet $q(\mathbf{k_2k_3})$ and creation of another quartet  $q(\mathbf{k_1k_2})$, where $\mathbf{k_2}$ is kept fixed while ${\cp k_3}$ is changed to ${\cp k_1}$. These diagrams contribute to
\begin{align}
\left[M_{11}(q) \right]_{\mathbf{k_2k_3};\mathbf{k_1k_2}}&=\frac{1}{2\beta}  \sum_{\omega_n}G_{11}^0(\mathbf{k_1} ,\mathbf{k_2},\mathbf{k_3},\mathrm{i}\omega_n)\nonumber\\&\times [G_{22}^0(\mathbf{k_1-q},\mathbf{k_2},\mathbf{k_3},\mathrm{i}\omega_n-\mathrm{i}v_n)\nonumber\\&+G_{22}^0(\mathbf{k_1},\mathbf{k_2},\mathbf{k_3-q},\mathrm{i}\omega_n-\mathrm{i}v_n)].\label{M11offdiag}
\end{align}
The prefactor $1/2$ above arises because the two terms contribute equally, corresponding to the free choice of the three momenta in $G_{22}^0$ (bold red line), i.e., the Green's function of the light fermion. This factor also ensures the continuity of $M(q)$ when $\mathbf{q}\to\mathbf{0}$. Similarly, as shown in Fig.\ref{fig_diagram}(d1,d2), $M_{21}$ also has elements that correspond to the annihilation of two quartets $q(\mathbf{k_2k_3})$ and $q(\mathbf{k_1k_2})$, which read
\begin{align}
\left[M_{21}(q) \right]_{\mathbf{k_2k_3};\mathbf{k_1k_2}}&=\frac{1}{2\beta} \sum_{\omega_n}G_{21}^0(\mathbf{k_1} ,\mathbf{k_2},\mathbf{k_3},\mathrm{i}\omega_n)\nonumber\\&\times[G_{21}^0(\mathbf{k_1+q},\mathbf{k_2},\mathbf{k_3},\mathrm{i}\omega_n+\mathrm{i}v_n)\nonumber\\&+G_{21}^0(\mathbf{k_1},\mathbf{k_2},\mathbf{k_3-q},\mathrm{i}\omega_n-\mathrm{i}v_n)].\label{M21offdiag}
\end{align}
The remaining matrices $M_{22}$ and $M_{12}$ can be obtained following the same diagrammatic rules, and we can verify that $M_{22}(q)=M_{11}(-q)$, $M_{12}(q)=M_{21}(-q)$ and $M(q)$ is a symmetric matrix.

The $M$ matrix defined above obeys the Goldstone theorem\cite{Randeria,QF2D0}, i.e.,
\begin{equation}
\det M(q=0)=0. \label{Goldstone}
\end{equation}
This is guaranteed by the mean-field gap equation (\ref{gapequation_T0}). Specifically, when $\mathbf{q=0}$ and $v_n=0$, $M_{11}(0)=M_{22}(0)=M_D$ and $M_{12}(0)=M_{21}(0)=M_{OD}$, so
\begin{equation}
    \det M(0)=\det (M_D+M_{OD})\det (M_{D}-M_{OD}).
\end{equation}
Since $G_{12}^0=G_{21}^0$, we can verify that $M_D-M_{OD}$ is exactly the matrix corresponding to the gap equation (further details are given in Appendixes \ref{sec:appendix_gap_equation} and  \ref{sec:appendix_math_M}); hence $\det (M_{D}-M_{OD})=0$, and consequently $\det M(0)=0$.

Finally, we obtain the thermodynamic potential induced by quantum fluctuations
\begin{equation}
\Omega_{\rm QF}=\frac{1}{2\beta}\sum_{q}\ln \det M. \label{Omega_QF}
\end{equation}
We note that $\Omega_{\rm QF}$ is formal and requires a proper regularization to yield a finite result. Following the regularization scheme similar to that used in the 2D pairing superfluid \cite{QF2D0}, we can obtain the regularized form of $\Omega_{\mathrm{QF}}$, as detailed in Appendix \ref{sec:appendixreg}.

In the following, we mainly focus on the deep binding limit where $\mu_Q\rightarrow E_4+0^+$, with $E_4\ (<0)$ being the quartet binding energy in vacuum. In this case each quartet cluster can be viewed as a composite boson, and we define a reduced chemical potential for the composite bosons
\begin{equation}
\tilde{\mu}_Q\equiv \mu_Q-E_4 \ (>0).\label{mu_Q}
\end{equation}
We will prove below that the leading order of  $\Omega_{\mathrm{QF}}$ in powers of $\tilde{\mu}_Q$ is given by
\begin{equation}
\Omega_{\rm QF}= \frac{SM_Q}{8\pi} \tilde{\mu}_Q^2 \ln \left( \frac{\tilde{\mu}_Q}{E_{2b}} + {\rm const}\right), \label{Omega_qf}
\end{equation}
where $M_Q=m_l+3m_h$ is the mass of a quartet. Importantly, Eq.(\ref{Omega_qf}) exhibits a logarithmic dependence on $\tilde{\mu}_Q$, well reproducing the leading term of EOS for 2D bosons\cite{2DBoson}. In this sense, it offers a correct physical picture of deeply bound  quartet clusters behaving as composite bosons. 

To prove Eq.(\ref{Omega_qf}), we first note that in the low-energy limit $\mathbf{q}\to0$ and $v_n \to 0$, $\det M \propto (E_q+\tilde{\mu}_Q)^2+v_n^2-\tilde{\mu}_Q^2$, where $E_q=q^2/2M_Q$. This expression is exactly proportional to the determinant of a $2\times 2$ matrix $\tilde{M}$,
\begin{equation}
    \tilde{M}\propto\begin{pmatrix} -\mathrm{i}v_n+E_q+\tilde{\mu}_Q   & \tilde{\mu}_Q \\ \tilde{\mu}_Q & \mathrm{i}v_n+E_q+\tilde{\mu}_Q\end{pmatrix},\label{bogoliubovboson}
\end{equation}
which is the inverse bosonic propagator in the Bogoliubov theory for repulsive bosons (a detailed derivation is given in Appendix \ref{sec:appendixM_structure}). It is then evident that the dominant contribution to $\Omega_{\rm QF}$ in Eq.(\ref{Omega_QF}) comes from the low-energy part, which can be reduced to the following integral
\begin{equation}
    I_{\epsilon}=-\frac{SM_Q}{4\pi^2}\int_{0}^{\epsilon} \mathrm{d}E_q\int_{0}^{\epsilon}\mathrm{d}\omega\frac{\tilde{\mu}_Q^2}{(E_q+\tilde{\mu}_Q)^2+\omega^2}.
\end{equation}
Evaluating this integral leads to the leading term in Eq.(\ref{Omega_qf}). In Appendix \ref{sec:appendixasy} we further show that the other terms of $\Omega_{\rm QF}$ are at most of order $\tilde{\mu}_Q^2$. Hence, we have analytically proved that the leading order of $\Omega_{\mathrm{QF}}$ in the deep binding limit is precisely given by Eq.(\ref{Omega_qf}).

\begin{figure}
    \centering
    \includegraphics[width=1\linewidth]{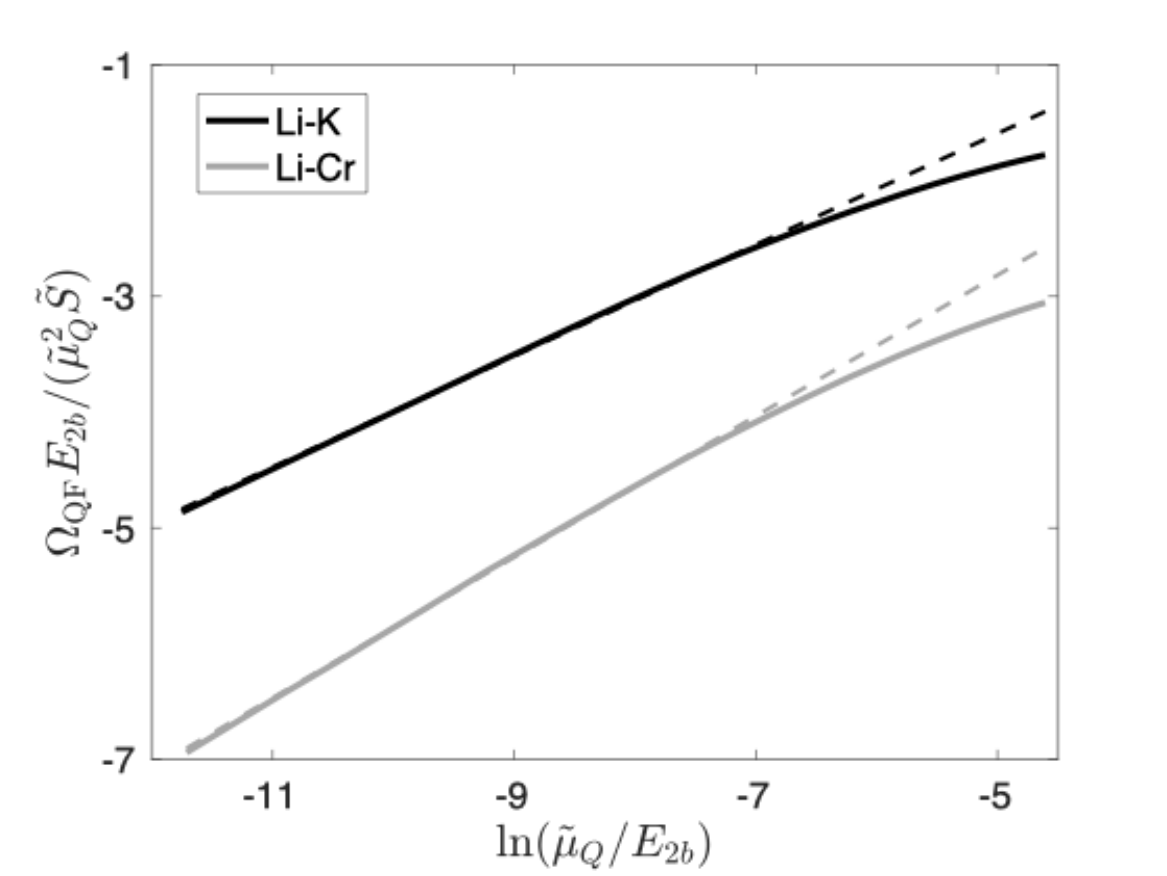}
    \caption{Thermodynamic potential of QSF from quantum fluctuations ($\Omega_{\mathrm{QF}}$) as a function of $\ln(\tilde{\mu}_Q/E_{2b})$, with $\tilde{\mu}_Q$ being the reduced chemical potential of quartet bosons [see definition in Eq.(\ref{mu_Q})]. The solid and dashed lines correspond to numerical results based on Eq.(\ref{Omega_QF}) and an analytical fit based on Eq.(\ref{Omega_qf}), respectively. The black and gray colors respectively denote $^{6}$Li-$^{40}$K and $^{6}$Li-$^{53}$Cr systems with different heavy-light mass ratios. Here we take the energy unit as $E_{2b}=1/(2m_ra^2)$, and denote $\tilde{S}=S/a^2$.}
    \label{fig_num}
\end{figure}

We have performed  numerical simulations of $\Omega_{\mathrm{QF}}$ based on Eq.(\ref{Omega_QF}) for two realistic mass-imbalanced mixtures of $^{6}$Li-$^{40}$K and $^{6}$Li-$^{53}$Cr. The results are shown as solid lines in Fig.\ref{fig_num}. Specifically, we plot $y\equiv \Omega_{\mathrm{QF}}E_{2b}/(\tilde{\mu}_Q^2 \tilde{S})$ ($\tilde{S}=S/a^2$) as a function of $x\equiv\ln(\tilde{\mu}_Q/E_{2b})$. In the deep binding limit ($\tilde{\mu}_Q\rightarrow 0^+$ and thus $x\rightarrow-\infty$), $y$ scales linearly with $x$ and the slope $y/x$ fits the analytical result $M_{Q}/(16\pi m_r)$ for both systems, as shown by dashed lines in Fig.\ref{fig_num}. This is consistent with the logarithmic dependence of $\Omega_{\mathrm{QF}}$ for QSF in the deep BEC limit. More details on the numerical method can be found in Appendix \ref{sec:appendixnum_cal}.

We note that the logarithmic dependence of $\Omega_{\rm QF}$  has been found previously in 2D pairing superfluids with Gaussian fluctuations\cite{QF2D0}, where an effective dimer-dimer scattering length was extracted from the total thermodynamic potential
\begin{equation}
\Omega=\Omega_0+\Omega_{\rm QF},
\end{equation}
by matching its form with the EOS of 2D bosons\cite{2DBoson}.  In principle, one can treat QSF similarly and extract an effective quartet-quartet scattering length. However, in our formalism, the mean-field part $\Omega_0$ scales as $S^3$ due to the neglect of the Pauli-blocking effect in treating the scatterings of triple momenta $\{{\mathbf{ k_1k_2k_3}}\}$. In contrast, the quantum fluctuation part $\Omega_{\mathrm{QF}}$ scales as $S$ and is therefore extensive. This is because externally Eq.(\ref{Omega_QF}) only involves summation over the center-of-mass momentum $\mathbf{q}$, with all internal momenta contained in the matrix $M$. For the total $\Omega/S$, although its leading order [$\sim M_Q\tilde{\mu}_Q^2\ln(\tilde{\mu}_Q/E_{2b})$] is well produced by quantum fluctuations, its next-leading order depends unphysically on the system area ($\sim S^2M_Q\tilde{\mu}_Q^2$). Since the quartet-quartet scattering length relies on the next-leading order in $\Omega/S$, we cannot arrive at a physical result for it at the moment. This is a shortcoming of our theory, which may be overcome in the future by employing other sophisticated methods, such as quantum Monte Carlo, to carefully incorporate the Pauli-blocking effect in evaluating $\Omega$. Nevertheless, the successful part of our theory is that it captures the dominant term $\sim M_Q\tilde{\mu}_Q^2\ln(\tilde{\mu}_Q/E_{2b})$ in the EOS, and during this process one can see clearly how the few-body correlation can be built up in Gaussian fluctuations of a high-order superfluid. 

At finite temperature, the quartet-quartet scattering length $a_{qq}$ in the deep binding limit determines the BKT transition temperature $T_{\mathrm{BKT}}$, which can be evaluated through the 2D quasicondensation of bosons \cite{BKTestimate} as $T_{\mathrm{BKT}}/T_{F}=1/\ln[-\xi \ln(k_F^2a_{qq}^2/4\pi)/4\pi]$, where $\xi \approx 380$ and $T_{F}=k_{F}^2/2M_{Q}$.

\section{Summary} \label{summary}

In summary, we have presented the Gaussian fluctuation theory for quartet superfluid (QSF) in 2D mass-imbalanced Fermi mixtures. Here QSF represents a high-order superfluid dominated by ($1+3$) cluster correlation that is well beyond the conventional two-body correlation in the BCS pairing framework. Based on a functional path integral approach that properly incorporates dominant four-body correlations in the quantum fluctuations of QSF, we have successfully reproduced the leading logarithmic contribution to the equation of state on the reduced chemical potential for quartet clusters, thereby offering a correct physical picture of these clusters behaving as composite bosons in the deep binding limit. Our results comprise attempt to evaluate quantum fluctuations in high-order fermion superfluids, which  shed light on  fluctuation effects in general fermion superfluids with arbitrarily high-order correlations. 

Future improvement of our work  involves a careful treatment of the Pauli-blocking effect when handling  triple momenta in representing a  quartet. This will allow the extraction of effective quartet-quartet coupling strength in the deep binding limit and, with it, the composite-boson theory for QSF can hopefully be completed.

Data that support the findings of this study are available \cite{Data}.

\bigskip

{\it Acknowledgements.} This work is supported by the National Natural Science Foundation of China (12525412, 92476104, 12134015) and the Quantum Science and Technology-National Science and Technology Major Project (2024ZD0300600).

\appendix
\setcounter{figure}{0}
\setcounter{equation}{0}
\renewcommand{\figurename}{Appendix Fig.}
\renewcommand{\thefigure}{\thesection\arabic{figure}}
\renewcommand{\theequation}{\thesection\arabic{equation}}
\setcounter{section}{0}

\section{Gap equation}\label{sec:appendix_gap_equation}

From the mean-field thermodynamic potential $\Omega_0=S_0/\beta$, we obtain the gap equation as
\begin{equation}
    -\frac{S}{g}\Delta_{\mathbf{k_2k_3}}+\frac{1}{\beta}\sum_{k_1}{'}\det G_{0}\Delta_{\mathbf{k_1k_2k_3}}=0. \label{Sgapequation}
\end{equation}
The above equation can be rewritten as $\det M_{\mathrm{GE}}=0$ and $M_{\mathrm{GE}}\mathbf{\Delta}=0$, where $\mathbf{\Delta}=\begin{pmatrix}...,\Delta_{\mathbf{k_2k_3} }, ... \end{pmatrix}^T$ and $M_{\mathrm{GE}}$ is a matrix with diagonal elements $-S/g+\sum_{k_1}{'}\det G_{0}/\beta$ and off-diagonal elements $\pm\sum_{\omega_n}\det G_0 /\beta$ or 0. The detailed construction of $M_{\mathrm{GE}}$ follows a similar procedure to that of $M_{ij}$, which is presented in Appendix \ref{sec:appendix_math_M}.

After performing the Matsubara summation, we obtain Eqs.(\ref{Omega}, \ref{gapequation}) in the main text. At $T=0$, the gap equation is given by
\begin{equation}
-\frac{S}{g} \Delta_{\mathbf{k_2k_3}}-\sum_{\mathbf{k_1}}\frac{\Delta_{\mathbf{k_1k_2k_3}}}{2\sqrt{\xi_{\mathbf{k_1k_2k_3}}^{+2}+\Delta_{\mathbf{k_1k_2k_3}}^2}}=0.\label{Sgapequation0}
\end{equation}
In the deep binding limit, the gap equation reduces exactly to the few-body equation in \cite{Cui1}, i.e.,
\begin{equation}
    -\frac{S\Delta_{\mathbf{k_2k_3}}}{g}-\sum_{\mathbf{k_1}}\frac{\Delta_{\mathbf{k_1k_2k_3}}}{\epsilon_{\mathbf{k_1k_2k_3}}-E_4}=0,\label{fewbodyequ}
\end{equation}
where $\epsilon_{\mathbf{k_1k_2k_3}} = \epsilon_{\mathbf{-k_1-k_2-k_3}}^l+\epsilon_{\mathbf{k_1}}^h +\epsilon_{\mathbf{k_2}}^h +\epsilon_{\mathbf{k_3}}^h$ and $E_4$ is the four-body binding energy. We can also rewrite Eq.(\ref{fewbodyequ}) as $\det M_{\mathrm{FE}}=0$ and $M_{\mathrm{FE}}\mathbf{\Delta}=0$ where $M_{\mathrm{FE}}$ is a matrix with diagonal elements $-S/g-\sum_{\mathbf{k}_1}{'}1/(\epsilon_{\mathbf{k_1k_2k_3}}-E_4)$ and off-diagonal elements $\mp 1/(\epsilon_{\mathbf{k_1k_2k_3}}-E_4)$ or 0. 

In the deep binding limit, we can expand the gap equation (\ref{Sgapequation0}) to third order in $\Delta$. Further utilizing the few-body equation in Eq.(\ref{fewbodyequ}), we get
\begin{equation}
    \sum_{\mathbf{k_1}}\frac{2\Delta_{\mathbf{k_1k_2k_3}}^3}{(\epsilon_{\mathbf{k_1k_2k_3}}-E_4)^3}=\sum_{\mathbf{k_1}}\frac{\Delta_{\mathbf{k_1k_2k_3}}}{(\epsilon_{\mathbf{k_1k_2k_3}}-E_4)^2}\tilde{\mu}_Q.\label{delta0Eq0}
\end{equation}
Given that  $\sqrt{\tilde{\mu}_Q}$ and $\tilde{\mu}_Q$ are both small ($\ll |E_4|$) in the deep binding limit, the above equation indicates that the amplitude of $\Delta_{\mathbf{k_2k_3}}$ scales as $\sqrt{\tilde{\mu}_Q}$ and approaches $0$ in this limit. 

In our numerical calculations, we find that for the ground state of the few-body and many-body system, these matrices $M_{\mathrm{GE}}$ and $M_{\mathrm{FE}}$ both have only one zero eigenvalue. This indicates that the ground state is nondegenerate. 

\section{Mathematical structure of inverse boson propagator $M$}\label{sec:appendix_math_M}
Similar to the pairing superfluid\cite{Melo2,HuHui1,Randeria,QF2D0,QF2DBKT,Salasnich}, we define the fluctuation matrix
\begin{widetext}
\begin{equation}
    W(\{\mathbf{k}\},\{\mathbf{k}'\},\mathrm{i}\omega_n,\mathrm{i}\omega_n')=\begin{pmatrix}
0  &\eta[(\{\mathbf{k}\},\{\mathbf{k}'\},\mathrm{i}v_n)]/\sqrt{\beta }  \\
 \bar{\eta}[(\{\mathbf{k}'\},\{\mathbf{k}\},-\mathrm{i}v_n)]/\sqrt{\beta } &0
\end{pmatrix},
\end{equation}
where the fluctuation field $\eta[(\{\mathbf{k}\},\{\mathbf{k}'\},\mathrm{i}v_n)]=\eta_{{\mathbf{k}_2}\mathbf{k}_3}(q)+\eta_{{\mathbf{k}_3}\mathbf{k}_1}(q)+\eta_{{\mathbf{k}_1}\mathbf{k}_2}(q)$, $\mathbf{q}=\mathbf{k}_1+\mathbf{k}_2+\mathbf{k}_3-\mathbf{k}_1'-\mathbf{k}_2'-\mathbf{k}_3'$ and $\mathrm{i}v_n=\mathrm{i}\omega_n-\mathrm{i}\omega_n'$. A given fluctuation term is nonzero only if the two momentum sets $\{\mathbf{k}\}$ and $\{\mathbf{k'}\}$ share at least two momenta in common. Then using $G^{-1}=G_0^{-1}+W$ and Eq.(\ref{effS}), the fluctuation action $S_2$ takes the form
\begin{equation}
    S_2=\frac{1}{2}\mathrm{tr}(G_0WG_0W)-\frac{S}{g}\sum_{q}\sum_{\{\mathbf{k}_2\mathbf{k}_3\}}\bar{\eta}_{\mathbf{k}_2\mathbf{k}_3}(q)\eta_{\mathbf{k}_2\mathbf{k}_3}(q),
\end{equation}
where
\begin{equation}
    \mathrm{tr}(G_0WG_0W)=\frac{1}{\beta}\sum_{\{\mathbf{k}\}}\sum_{\{\mathbf{k}'\}}\sum_{\omega_n,v_n}\mathrm{F}^{\dagger}\begin{pmatrix}
G_{11}^0(\{\mathbf{k}\},\mathrm{i}\omega_n)G_{22}^0(\{\mathbf{k}'\},\mathrm{i}\omega_n')  &G_{12}^0(\{\mathbf{k}\},\mathrm{i}\omega_n)G_{12}^0(\{\mathbf{k}'\},\mathrm{i}\omega_n')  \\
 G_{21}^0(\{\mathbf{k}\},\mathrm{i}\omega_n)G_{21}^0(\{\mathbf{k}'\},\mathrm{i}\omega_n') & G_{22}^0(\{\mathbf{k}\},\mathrm{i}\omega_n)G_{11}^0(\{\mathbf{k}'\},\mathrm{i}\omega_n')
\end{pmatrix}\mathrm{F}
\end{equation}
and $\mathrm{F}=(\eta[(\{\mathbf{k}\},\{\mathbf{k}'\},\mathrm{i}v_n)],\eta[(\{\mathbf{k}'\},\{\mathbf{k}\},-\mathrm{i}v_n)])^T$.

We then consider $M_{11}$, which is derived from this term $G_{11}^0(\{\mathbf{k}\},\mathrm{i}\omega_n)G_{22}^0(\{\mathbf{k}'\},\mathrm{i}\omega_n')\bar{\eta}[(\{\mathbf{k}\},\{\mathbf{k}'\},\mathrm{i}v_n)]\eta[(\{\mathbf{k}\},\{\mathbf{k}'\},\mathrm{i}v_n)]$. For $\mathbf{q}=0$, expanding the fluctuation field and using symmetry and cyclic invariance, we obtain three terms:
\begin{align}
\sum_{\{\mathbf{k}_2\mathbf{k}_3\}} \sum_{k_1}{'}\sum_{v_n}G_{11}^0(\{\mathbf{k}\},\mathrm{i}\omega_n)G_{22}^0(\{\mathbf{k}\},\mathrm{i}\omega_n-\mathrm{i}v_n)\bar{\eta}_{\mathbf{k}_2\mathbf{k}_3}(\mathbf{0},\mathrm{i}v_n)[\eta_{\mathbf{k}_2\mathbf{k}_3}(\mathbf{0},\mathrm{i}v_n)+\eta_{\mathbf{k}_3\mathbf{k}_1}(\mathbf{0},\mathrm{i}v_n)+\eta_{\mathbf{k}_1\mathbf{k}_2}(\mathbf{0},\mathrm{i}v_n)]
\end{align}
For a given $\mathbf{q}\ne \mathbf{0}$, we obtain
\begin{align}
    &\sum_{\{\mathbf{k}\}}\sum_{\omega_n,v_n}G_{11}^0(\{\mathbf{k}\},\mathrm{i}\omega_n)G_{22}^0(\{\mathbf{k}'\},\mathrm{i}\omega_n-\mathrm{i}v_n)\bar{\eta}[(\{\mathbf{k}\},\{\mathbf{k}'\},\mathrm{i}v_n)]\eta[(\{\mathbf{k}\},\{\mathbf{k}'\},\mathrm{i}v_n)]\nonumber\\=&\sum_{\{\mathbf{k}\}}\sum_{\omega_n,v_n}G_{11}^0(\mathbf{k}_1,\mathbf{k}_2,\mathbf{k}_3,\mathrm{i}\omega_n)G_{22}^0(\mathbf{k}_1-\mathbf{q},\mathbf{k}_2,\mathbf{k}_3,\mathrm{i}\omega_n-\mathrm{i}v_n)\bar{\eta}[(\{\mathbf{k}\},\{\mathbf{k}'\},\mathrm{i}v_n)]\eta[(\{\mathbf{k}\},\{\mathbf{k}'\},\mathrm{i}v_n)]\nonumber\\+&\sum_{\{\mathbf{k}\}}\sum_{\omega_n,v_n}G_{11}^0(\mathbf{k}_1,\mathbf{k}_2,\mathbf{k}_3,\mathrm{i}\omega_n)G_{22}^0(\mathbf{k}_1,\mathbf{k}_2-\mathbf{q},\mathbf{k}_3,\mathrm{i}\omega_n-\mathrm{i}v_n)\bar{\eta}[(\{\mathbf{k}\},\{\mathbf{k}'\},\mathrm{i}v_n)]\eta[(\{\mathbf{k}\},\{\mathbf{k}'\},\mathrm{i}v_n)]\nonumber\\+&\sum_{\{\mathbf{k}\}}\sum_{\omega_n,v_n}G_{11}^0(\mathbf{k}_1,\mathbf{k}_2,\mathbf{k}_3,\mathrm{i}\omega_n)G_{22}^0(\mathbf{k}_1,\mathbf{k}_2,\mathbf{k}_3-\mathbf{q},\mathrm{i}\omega_n-\mathrm{i}v_n)\bar{\eta}[(\{\mathbf{k}\},\{\mathbf{k}'\},\mathrm{i}v_n)]\eta[(\{\mathbf{k}\},\{\mathbf{k}'\},\mathrm{i}v_n)],
\end{align}
which gives nine terms after expanding the fluctuation field. Among these nine terms, some are unphysical. For example, we will obtain the terms like $G_{22}^0(\mathbf{k_1}-\mathbf{q},\mathbf{k}_2,\mathbf{k}_3,\mathrm{i}\omega_n-\mathrm{i}v_n)G_{11}^0(\mathbf{k_1},\mathbf{k}_2,\mathbf{k}_3,\mathrm{i}\omega_n-\mathrm{i}v_n)\bar{\eta}_{\mathbf{k}_3\mathbf{k}_1}(q)\eta_{\mathbf{k}_3\mathbf{k}_1}(q)$, where $\mathbf{k}_1$ should remain unchanged according to the fluctuation field, yet in $G_{22}^0$ it is shifted. Such terms should be excluded. After this exclusion, only five terms remain, i.e. 
\begin{align}
    &\sum_{\{\mathbf{k}_2\mathbf{k}_3\}}\sum_{k_1}{'}\sum_{v_n}G_{11}^0(\{\mathbf{k}\},\mathrm{i}\omega_n)G_{22}^0(\mathbf{k}_1-\mathbf{q},\mathbf{k}_2,\mathbf{k}_3,\mathrm{i}\omega_n-\mathrm{i}v_n)\bar{\eta}_{\mathbf{k}_2\mathbf{k}_3}(q)\eta_{\mathbf{k}_2\mathbf{k}_3}(q)\nonumber\\+&\sum_{\{\mathbf{k}_2\mathbf{k}_3\}}\sum_{k_1}{'}\sum_{v_n}G_{11}^0(\{\mathbf{k}\},\mathrm{i}\omega_n)G_{22}^0(\mathbf{k}_1-\mathbf{q},\mathbf{k}_2,\mathbf{k}_3,\mathrm{i}\omega_n-\mathrm{i}v_n)\bar{\eta}_{\mathbf{k}_2\mathbf{k}_3}(q)\eta_{\mathbf{k}_3\mathbf{k}_1}(q)\nonumber\\+&\sum_{\{\mathbf{k}_2\mathbf{k}_3\}}\sum_{k_1}{'}\sum_{v_n}G_{11}^0(\{\mathbf{k}\},\mathrm{i}\omega_n)G_{22}^0(\mathbf{k}_1-\mathbf{q},\mathbf{k}_2,\mathbf{k}_3,\mathrm{i}\omega_n-\mathrm{i}v_n)\bar{\eta}_{\mathbf{k}_2\mathbf{k}_3}(q)\eta_{\mathbf{k}_1\mathbf{k}_2}(q)\nonumber\\+&\sum_{\{\mathbf{k}_2\mathbf{k}_3\}}\sum_{k_1}{'}\sum_{v_n}G_{11}^0(\{\mathbf{k}\},\mathrm{i}\omega_n)G_{22}^0(\mathbf{k}_1,\mathbf{k}_2,\mathbf{k}_3-\mathbf{q},\mathrm{i}\omega_n-\mathrm{i}v_n)\bar{\eta}_{\mathbf{k}_2\mathbf{k}_3}(q)\eta_{\mathbf{k}_1\mathbf{k}_2}(q)\nonumber\\+&\sum_{\{\mathbf{k}_2\mathbf{k}_3\}}\sum_{k_1}{'}\sum_{v_n}G_{11}^0(\{\mathbf{k}\},\mathrm{i}\omega_n)G_{22}^0(\mathbf{k}_1,\mathbf{k}_2-\mathbf{q},\mathbf{k}_3,\mathrm{i}\omega_n-\mathrm{i}v_n)\bar{\eta}_{\mathbf{k}_2\mathbf{k}_3}(q)\eta_{\mathbf{k}_3\mathbf{k}_1}(q).\label{ExpandTerm}
\end{align}
\end{widetext} 
The first term in Eq.(\ref{ExpandTerm}) corresponds to Fig.\ref{fig_diagram}(b1), i.e., the diagonal part of $M_{11}(q)$, while the remaining four correspond to the off-diagonal elements, with the third and fourth terms corresponding to Fig.\ref{fig_diagram}(c1,c2). However, due to the free choice of the three momenta in $G_{22}^0$, $M(q)$ is not continuous between the $\mathbf{q}=0$ and $\mathbf{q}\to\mathbf{0}$ limit. To restore this continuity, and given that the two contributions are equally weighted, we introduce an extra factor $1/2$. This leads to Eq.(\ref{M11offdiag}) in the main text. The above procedure can be similarly applied to obtain $M_{22}$, $M_{12}$ and $M_{21}$.  

To express the fluctuation terms in matrix form, we first define the basis in $\left\{\mathbf{k}_i\mathbf{k}_j\right\}$ space. We assign each of the $N_k$ discrete momenta a unique index $\mathrm{Ind}(\mathbf{k})$, and order each pair $\left\{\mathbf{k}_i\mathbf{k}_j\right\}$ by the rule $\mathrm{Ind}(\mathbf{k}_i)<\mathrm{Ind}(\mathbf{k}_j)$. This defines a set of $N_k(N_k-1)/2$ basis vectors. With this basis, the fluctuation fields $\eta(q)$ and $\bar{\eta}(q)$ in the main text are $N_k(N_k-1)/2$-dimensional column vectors, and each $M_{ij}(q)$ is an $N_k(N_k-1)/2\times N_k(N_k-1)/2$ matrix.

For a given row labeled by $\left\{\mathbf{k}_i\mathbf{k}_j\right\}$ of $M_{ij}$, the nonzero elements occur in columns labeled by $\left\{\mathbf{k}\mathbf{k}_i\right\}$ or $\left\{\mathbf{k}\mathbf{k}_j\right\}$. Thus each row has $2N_k-3$ nonzero entries. The sign convention is as follows: for column $\left\{\mathbf{k}\mathbf{k}_i\right\}$, the sign is positive if $\mathrm{Ind}(\mathbf{k})<\mathrm{Ind}(\mathbf{k}_i)$ and negative if $\mathrm{Ind}(\mathbf{k})>\mathrm{Ind}(\mathbf{k}_i)$; for column $\left\{\mathbf{k}\mathbf{k}_j\right\}$, the sign is positive if $\mathrm{Ind}(\mathbf{k})>\mathrm{Ind}(\mathbf{k}_j)$ and negative if $\mathrm{Ind}(\mathbf{k})<\mathrm{Ind}(\mathbf{k}_j)$. With these rules, we can verify $M_{22}(q)=M_{11}(-q)$, $M_{12}(q)=M_{21}(-q)$ and $M(q)$ is a symmetric matrix.

We now prove the Goldstone theorem. At $q=0$, $M_{11}(0)=M_{22}(0)=M_D$ and $M_{12}(0)=M_{21}(0)=M_{OD}$. Then the determinant $\det M(0)=\det (M_D+M_{OD})\det (M_{D}-M_{OD})$. Because the diagonal and off-diagonal elements of $M_D$ are $-S/g+\sum_{k_1}'G_{11}^0G_{22}^0/\beta$ and $\pm \sum_{\omega_n}G_{11}^0G_{22}^0/\beta$ (or $0$), and those of $M_{OD}$ are $\sum_{k_1}'G_{12}^0G_{21}^0/\beta$ and $\pm \sum_{\omega_n}G_{12}^0G_{21}^0/\beta$ (or $0$), we can verify that $M_D-M_{OD}$ is exactly $M_{\mathrm{GE}}$ defined in Appendix \ref{sec:appendix_gap_equation}, which corresponds to the mean-field gap equation. Hence $\det (M_D-M_{OD})=0$, which implies $\det M(0)=0$, and the null eigenvector of $M_D-M_{OD}$ is precisely the solution of the gap equation $\mathbf{\Delta}=\begin{pmatrix}...,\Delta_{\mathbf{k_2k_3} }, ... \end{pmatrix}^T$. 

Moreover, for any nonzero vector $\mathbf{f}=(...,f_{\mathbf{k_2k_3}},...)$, we obtain
\begin{equation}
    \mathbf{f}^TM_{OD}\mathbf{f}=\frac{1}{\beta}\sum_{\mathbf{\left\{k\right\}}}G_{12}^0G_{21}^0(f_{\mathbf{k}_2\mathbf{k}_3}+f_{\mathbf{k}_3\mathbf{k}_1}+f_{\mathbf{k}_1\mathbf{k}_2})^2>0,
\end{equation}
so $M_{OD}$ is positive definite.  Thus $\det(M_D+M_{OD})>0 $, meaning that the zero mode of $M(0)$ is unique. This confirms that the Goldstone mode is nondegenerate.

\section{Structure of inverse boson propagator $M$ in the deep binding  limit}
\label{sec:appendixM_structure}

We first perform the summation over the Matsubara frequencies $\omega_n$ in $M(q)$. At $T=0$, we have  
\begin{widetext}
\begin{align}
 \frac{1}{\beta}\sum_{\omega_n}G_{11}^0(\mathbf{k_1},\mathbf{k_2},\mathbf{k_3},\mathrm{i}\omega_n)G_{22}^0(\mathbf{k_1+q},\mathbf{k_2},\mathbf{k_3},\mathrm{i}\omega_n+\mathrm{i}v_n)&= \frac{v_{\mathbf{k_1k_2k_3}}^2v_{\mathbf{k_1+q,k_2k_3}}^2}{\mathrm{i}v_n+E_{{\mathbf{k_1k_2k_3}}}^--E_{\mathbf{k_1+q,k_2k_3}}^+}-\frac{u_{\mathbf{k_1k_2k_3}}^2u_{\mathbf{k_1+q,k_2k_3}}^2}{\mathrm{i}v_n+E_{{\mathbf{k_1k_2k_3}}}^+-E_{\mathbf{k_1+q,k_2k_3}}^-},\nonumber \\ \frac{1}{\beta}\sum_{\omega_n}G_{11}^0(\mathbf{k_1},\mathbf{k_2},\mathbf{k_3},\mathrm{i}\omega_n)G_{22}^0(\mathbf{k_1-q},\mathbf{k_2},\mathbf{k_3},\mathrm{i}\omega_n-\mathrm{i}v_n)&= \frac{u_{\mathbf{k_1k_2k_3}}^2u_{\mathbf{k_1-q,k_2k_3}}^2}{\mathrm{i}v_n-E_{{\mathbf{k_1k_2k_3}}}^++E_{\mathbf{k_1-q,k_2k_3}}^-} -\frac{v_{\mathbf{k_1k_2k_3}}^2v_{\mathbf{k_1-q,k_2k_3}}^2}{\mathrm{i}v_n-E_{{\mathbf{k_1k_2k_3}}}^-+E_{\mathbf{k_1-q,k_2k_3}}^+},\nonumber\\
\frac{1}{\beta}\sum_{\omega_n}G_{12}^0(\mathbf{k_1},\mathbf{k_2},\mathbf{k_3},\mathrm{i}\omega_n)G_{12}^0(\mathbf{k_1+q},\mathbf{k_2},\mathbf{k_3},\mathrm{i}\omega_n+\mathrm{i}v_n)&=\frac{\Delta
_{{\mathbf{k_1k_2k_3}}}\Delta_{\mathbf{k_1+q,k_2k_3}}}{4E_{{\mathbf{k_1k_2k_3}}}E_{\mathbf{k_1+q,k_2k_3}}}  [ \frac{1}{\mathrm{i}v_n+E_{{\mathbf{k_1k_2k_3}}}^+-E_{\mathbf{k_1+q,k_2k_3}}^-}\nonumber\\&-\frac{1}{\mathrm{i}v_n+E_{{\mathbf{k_1k_2k_3}}}^--E_{\mathbf{k_1+q,k_2k_3}}^+}], \nonumber\\ \frac{1}{\beta}\sum_{\omega_n}G_{12}^0(\mathbf{k_1},\mathbf{k_2},\mathbf{k_3},\mathrm{i}\omega_n)G_{12}^0(\mathbf{k_1-q},\mathbf{k_2},\mathbf{k_3},\mathrm{i}\omega_n-\mathrm{i}v_n)&=\frac{\Delta
_{{\mathbf{k_1k_2k_3}}}\Delta_{\mathbf{k_1-q,k_2k_3}}}{4E_{{\mathbf{k_1k_2k_3}}}E_{\mathbf{k_1-q,k_2k_3}}}  [ \frac{1}{\mathrm{i}v_n\nonumber-E_{{\mathbf{k_1k_2k_3}}}^-+E_{\mathbf{k_1-q,k_2k_3}}^+}\nonumber\\&-\frac{1}{\mathrm{i}v_n-E_{{\mathbf{k_1k_2k_3}}}^++E_{\mathbf{k_1-q,k_2k_3}}^-}],
\end{align}
\end{widetext}
where $E_{\mathbf{k_1k_2k_3}}= \sqrt{\xi_{\mathbf{k_1k_2k_3}}^{+2}+\Delta_{\mathbf{k_1k_2k_3}}^2 }$, $u_{\mathbf{k_1k_2k_3}}^2=(1+\xi_{\mathbf{k_1k_2k_3}}^{+}/\sqrt{\xi_{\mathbf{k_1k_2k_3}}^{+2}+\Delta_{\mathbf{k_1k_2k_3}}^2})/2$, and $v_{\mathbf{k_1k_2k_3}}^2=1-u_{\mathbf{k_1k_2k_3}}^2$.

In the deep binding limit $\tilde{\mu}_Q \to 0^+$, we have $u_{{\mathbf{k_1k_2k_3}}}^2 \approx1-\Delta_{{\mathbf{k_1k_2k_3}}}^2/4\xi_{{\mathbf{k_1k_2k_3}}}^{+2}$ and $v_{{\mathbf{k_1k_2k_3}}}^2\approx \Delta_{{\mathbf{k_1k_2k_3}}}^2/4\xi_{{\mathbf{k_1k_2k_3}}}^{+2}$, so nonzero matrix elements in $M_{11}$ and $M_{22}$ are dominated by  $u_{{\mathbf{k_1k_2k_3}}}^2u_{\{\mathbf{k_1\pm q}\}}^2$ terms. Therefore, we can decompose $M_{11}$ and $M_{22}$ into two parts: $M_{11}=M_{11}^c+M_{11}^r$ and $M_{22}=M_{22}^c+M_{22}^r$, where $M_{ii}^c$ contains $u_{{\mathbf{k_1k_2k_3}}}^2u_{\{\mathbf{k_1\pm q}\}}^2$ terms while $M_{ii}^r$ contains $v_{{\mathbf{k_1k_2k_3}}}^2v_{\{\mathbf{k_1\pm q}\}}^2$ terms. In this way, we separate $M$ into three parts: $M=M^c_{\mathrm{DB}}+M^r_{\mathrm{DB}}+M_{\mathrm{ODB}}$, with $M_{\mathrm{DB}}^c=\begin{pmatrix} M_{11}^c & 0 \\ 0 & M_{22}^c\end{pmatrix}$, $M_{\mathrm{DB}}^r= \begin{pmatrix} M_{11}^r & 0 \\ 0 & M_{22}^r\end{pmatrix}$, and $M_{\mathrm{ODB}}= \begin{pmatrix} 0 & M_{12} \\ M_{21} & 0\end{pmatrix}$. Moreover, in the deep binding limit, since $\Delta_{\mathbf{k_2k_3}}\propto \sqrt{\tilde{\mu}_Q}$, $M_{11}^r$ and $M_{22}^r$ are proportional to $\tilde{\mu}_Q^2$, while $M_{12}$ and $M_{21}$ are proportional to $\tilde{\mu}_Q$. 

According to the decomposition, we have 
\begin{equation}
    \det M=\det M_{\mathrm{DB}}^c\det[1+(M_{\mathrm{DB}}^{c})^{-1}(M_{\mathrm{DB}}^r+M_{\mathrm{ODB}})].
\end{equation}

We first consider $\det M_{\mathrm{DB}}^c$. According to the definition of $M_{\mathrm{DB}}^c$, one has $\det M_{\mathrm{DB}}^c=\det M_{11}^c\det M_{22}^c$. Then in the deep binding limit and the low-energy limit, we perform the Taylor expansion of $M_{11}^c (\mathrm{i}v_n,\tilde{\mu}_Q,q)$ and $M_{22}^c (\mathrm{i}v_n,\tilde{\mu}_Q,q)$, yielding
\begin{align}
    M_{11}^c&=A-\mathrm{i}v_nB+\tilde{\mu}_Q C-qD+q^2F, \\M_{22}^c &=A+\mathrm{i}v_nB+\tilde{\mu}_Q C+qD+q^2F.
\end{align}
It is straightforward to verify that the matrix $A$ precisely gives the matrix $M_{\rm FE}$ associated with the few-body equation. Thus, $\det M_{11}^c(0,0,0)=\det M_{22}^c(0,0,0)=\det A=0$, and we can obtain $A\mathbf{\Delta}=0$, where $\mathbf{\Delta}=\begin{pmatrix}...,\Delta_{\mathbf{k_2k_3} }, ... \end{pmatrix}^T$. Since the zero mode of $A$ is nondegenerate, we can introduce the adjugate matrix $\tilde{A}=\mathrm{adj}A$ of $A$ and obtain $\tilde{A}\propto \mathbf{\Delta\Delta}^T$. For convenience, we will denote $\tilde{A}=C_A\mathbf{\Delta\Delta}^T$ in the following.

Next, we perform the Taylor expansion of $\det M_{11}^c$ and $\det M_{22}^c$ up to the first order of $\tilde{\mu}_Q$, the first order of $v_n$ and the second order of $q$. Using
\begin{equation}
\det Z(x)\approx \det Z_0 +x\mathrm{tr}[\mathrm{adj(Z_0)}\frac{\mathrm{d}Z(x)}{\mathrm{d}x} |_{x=0}],
\end{equation}
where $Z_0=Z(0)$ and $\mathrm{adj}Z_0$ is the adjugate matrix of $Z_0$, we get 
\begin{align}
    \det M_{11}^c&=-q\mathrm{tr}(\tilde{A}D)+q^2[\mathrm{tr}(\tilde{A}F)+C_q^2]\nonumber\\&+\tilde{\mu}_Q\mathrm{tr}(\tilde{A}C)-\mathrm{i}v_n\mathrm{tr}(\tilde{A}B), \\ \det M_{22}^c&=q\mathrm{tr}(\tilde{A}D)+q^2[\mathrm{tr}(\tilde{A}F)+C_q^2]\nonumber\\&+\tilde{\mu}_Q\mathrm{tr}(\tilde{A}C)+\mathrm{i}v_n\mathrm{tr}(\tilde{A}B).
\end{align} 
Next we will demonstrate three properties:

(1) $\mathrm{tr}(\tilde{A}B) = \mathrm{tr}(\tilde{A}C)$

First, we have
\begin{align}
    \mathrm{tr}(\tilde{A}B)&=C_A\sum_{\left\{\mathbf{k}\right\}}\frac{\Delta_{{\mathbf{k_1k_2k_3}}}^2}{(\epsilon_{\mathbf{k_1k_2k_3}}-E_4)^2}, \label{dmu} \\
    \mathrm{tr}(\tilde{A}C)&=C_A\sum_{\left\{\mathbf{k}\right\}}\frac{4\Delta_{{\mathbf{k_1k_2k_3}}}^4}{(\epsilon_{\mathbf{k_1k_2k_3}}-E_4)^3\tilde{\mu}_Q}\nonumber\\&-C_A\sum_{\{\mathbf{k}\}}\frac{\Delta_{\mathbf{k_1k_2k_3}}^2}{(\epsilon_{\mathbf{k_1k_2k_3}}-E_4)^2}.
\end{align}
Recalling Eq.(\ref{delta0Eq0}), we then get $\mathrm{tr}(\tilde{A}B) = \mathrm{tr}(\tilde{A}C)$. 

(2) $\mathrm{tr}(\tilde{A}D)  = 0$

First, we have
\begin{align}
    \mathrm{tr}(\tilde{A}D)= C_A\sum_{\left\{ \mathbf{k}\right\}}\frac{\Delta_{{\mathbf{k_1k_2k_3}}}^2 |\mathbf{k}_{123}|}{(\epsilon_{\mathbf{k_1k_2k_3}}-E_4)^2}\frac{\cos (\theta_q-\theta_{123})}{m_l},
\end{align}
where $|\mathbf{k}_{123}|=|\mathbf{k}_1+\mathbf{k}_2+\mathbf{k}_3|$, and $\theta_q$ ($\theta_{123}$) is the angle of $\mathbf{q}$ ($\mathbf{k}_1+\mathbf{k}_2+\mathbf{k}_3$). We then define the summand as a new function $F(\mathbf{k}_1,\mathbf{k}_2,\mathbf{k}_3,\theta_q)$. Utilizing the rotational invariance of the gap, i.e., $\Delta_{\mathbf{k}_2\mathbf{k}_3}=\Delta_{-\mathbf{k}_2,-\mathbf{k}_3}$, we can conclude that $F(-\mathbf{k}_1,-\mathbf{k}_2,-\mathbf{k}_3,\theta_q)=-F(\mathbf{k}_1,\mathbf{k}_2,\mathbf{k}_3,\theta_q)$. This yields $\mathrm{tr}(\tilde{A}D)=0$.

(3) The coefficient of the $q^2$ term is $\mathrm{tr}(\tilde{A}B)/2M_Q$.

To prove this, we first introduce the generalized inverse $A^+$ satisfying $AA^+A=A$. Then using $\mathrm{tr}(\tilde{A}D)=0$, we obtain the expansion of $\det(A+qD+q^2F)$ to second order in $q$ as 
\begin{equation}
    \det(A+qD+q^2F) \approx q^2[\mathrm{tr}(\tilde{A}F)-\mathrm{tr}(\tilde{A}DA^{+}D)].
\end{equation}
Then we consider the few-body equation with a macroscopic momentum $\mathbf{q}$, i.e., 
\begin{align}
    &-\frac{S}{g}\Delta_{\mathbf{k_2k_3}}(\mathbf{q})\nonumber\\&=\sum_{\mathbf{k_1}}\frac{\Delta_{{\mathbf{k_1k_2k_3}}}(\mathbf{q})}{\epsilon_{\mathbf{q-k_1-k_2-k_3}}^l+\epsilon_{\mathbf{k_1}}^h+\epsilon_{\mathbf{k_2}}^h +\epsilon_{\mathbf{k_3}}^h-E}, \label{qfewbodyequ}
\end{align}
where $E=E_4+E_q$ and $E_q=q^2/2M_Q$. Based on Eq.(\ref{qfewbodyequ}), we can similarly construct a new matrix $M_{\mathrm{FE}}(\mathbf{q})$, with $\det M_{\mathrm{FE}}(\mathbf{q})=0$ and $M_{\mathrm{FE}}(\mathbf{q})\mathbf{\Delta}(\mathbf{q})=0$. For a small $\mathbf{q}$, expand $M_{\mathrm{FE}}(\mathbf{q})$ and $\mathbf{\Delta}(\mathbf{q})$ as $M_{\mathrm{FE}}(\mathbf{q})=A+qD+q^2F-q^2B/2M_Q$ and $\mathbf{\Delta}(\mathbf{q})=\mathbf{\Delta}+q\mathbf{\Delta}_1+q^2\mathbf{\Delta}_2$. Substituting these expansions into $M_{\mathrm{FE}}(\mathbf{q})\mathbf{\Delta}(\mathbf{q})=0$ and comparing coefficients order by order yields (i) $A\mathbf{\Delta}=0$; (ii) $A\mathbf{\Delta}_1+D\mathbf{\Delta}=0$, which leads to $\mathbf{\Delta}_1=-A^+D\mathbf{\Delta}+\alpha \mathbf{\Delta}$ where $\alpha$ is arbitrary; (iii) $A\mathbf{\Delta}_2+D\mathbf{\Delta}_1+F\mathbf{\Delta}-B\mathbf{\Delta}/2M_Q=0$. Multiplying the last equation from the left by $\mathbf{\Delta}^T$, and using $\mathbf{\Delta}^TA=0$ and $\mathbf{\Delta}^TD\mathbf{\Delta}=0$, we obtain $\mathbf{\Delta}^TB\mathbf{\Delta}/2M_Q=\mathbf{\Delta}^TF\mathbf{\Delta}-\mathbf{\Delta}^TDA^{+}D\mathbf{\Delta}$. Finally, using $\tilde{A}=C_A\mathbf{\Delta}\mathbf{\Delta}^T$, we arrive at $\det(A+qD+q^2F)=\mathrm{tr}(\tilde{A}B)q^2/2M_Q$.

Based on the above properties, we finally obtain the expansions as 
\begin{align}
    &\det M_{11}^c=\mathrm{tr}(\tilde{A}B)(-\mathrm{i}v_n+E_q+\tilde{\mu}_Q), \label{detMc11}\\
    &\det M_{22}^c=\mathrm{tr}(\tilde{A}B)(\mathrm{i}v_n+E_q+\tilde{\mu}_Q). \label{detMc22}
\end{align}

Now we consider $\det[1+(M_{\mathrm{DB}}^{c})^{-1}(M_{\mathrm{DB}}^r+M_{\mathrm{ODB}})]$. As $\tilde{\mu}_Q \to 0$, $M_{\mathrm{DB}}^r\sim \tilde{\mu}_Q^2$, $M_{\mathrm{ODB}}\sim \tilde{\mu}_Q$, while only $M_{\mathrm{DB}}^c$ approaches a constant. Therefore, $(M_{\mathrm{DB}}^{c})^{-1}(M_{\mathrm{DB}}^r+M_{\mathrm{ODB}})$ is a small perturbation matrix. Applying the determinant expansion formula,
\begin{align}
    \det (1+xB)\approx1+x\mathrm{tr}B+\frac{x^2}{2}[(\mathrm{tr}B)^2-\mathrm{tr}B^2],
\end{align}
and $\mathrm{tr}[(M_{\mathrm{DB}}^{c})^{-1}M_{\mathrm{ODB}}]=0$, we obtain
\begin{align}
     &\det [1+(M_{\mathrm{DB}}^{c})^{-1}(M_{\mathrm{DB}}^r+M_{\mathrm{ODB}})] \nonumber\\
     &\approx 1+\mathrm{tr}[(M_{\mathrm{DB}}^{c})^{-1}M_{\mathrm{DB}}^r]\nonumber\\&-\frac{1}{2}\mathrm{tr}[(M_{\mathrm{DB}}^{c})^{-1}M_{\mathrm{ODB}}(M_{\mathrm{DB}}^{c})^{-1}M_{\mathrm{ODB}}].\label{detconvergent}
\end{align}

 Then we neglect the contribution of $M_{\mathrm{DB}}^r$ and expand $M_{\mathrm{ODB}}$ in the low-energy limit to the lowest order of $q$ and $v_n$. Using $M_{12}(0)=M_{21}(0)=M_{12}^0$ and $(M_{\mathrm{11}}^{c})^{-1}=\mathrm{adj}(M_{\mathrm{11}}^c)/\det M_{\mathrm{11}}^c$, we further expand $\mathrm{adj}(M_{\mathrm{11}}^c)$ to the lowest order, i.e., $\tilde{A}=\mathrm{adj}(A)$ and obtain 
 \begin{equation}
    \det M\approx \det M_{11}^c \det M_{22}^c-[\mathrm{tr}(\tilde{A}M_{12}^0)]^2,
\end{equation}
Recalling Eq.(\ref{delta0Eq0}), we can prove $\mathrm{tr}(\tilde{A}M_{12}^0)=\tilde{\mu}_Q\mathrm{tr}(\tilde{A}B)$, and therefore $\det M=[\mathrm{tr}(\tilde{A}B)]^2 [(q^2/2M_Q+\tilde{\mu}_Q)^2+v_n^2-\tilde{\mu}_Q^2]$. This expression is exactly the determinant of a $2\times2$ matrix $\tilde{M}$
\begin{equation}
    \tilde{M}=\mathrm{tr}(\tilde{A}B)\begin{pmatrix} -\mathrm{i}v_n+E_q+\tilde{\mu}_Q   & \tilde{\mu}_Q \\ \tilde{\mu}_Q & \mathrm{i}v_n+E_q+\tilde{\mu}_Q\end{pmatrix}.
\end{equation}
This form is identical to the inverse boson propagator in the Bogoliubov theory for repulsive bosons, indicating that in the deep binding and low-energy limits, our theory recovers the Bogoliubov analysis of repulsive bosons.

\section{Regularization of $\Omega_{\rm QF}$}\label{sec:appendixreg}

\begin{figure}[h]
\includegraphics[width=0.25\textwidth]{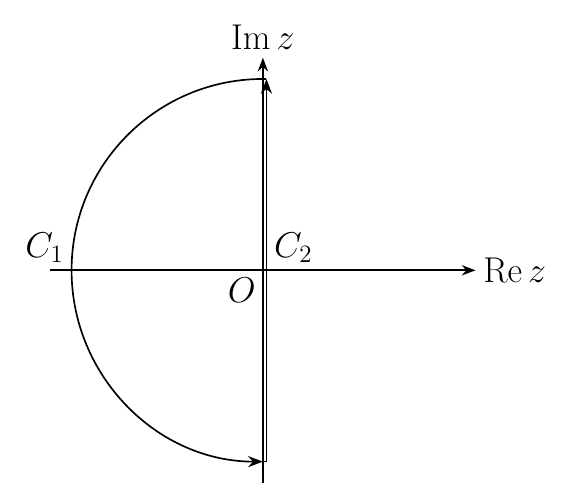}
\caption{ The integration contour consists of two parts, $C_1$ and $C_2$. Here, $C_1$ is a semicircle in the left half-plane, while $C_2$ runs parallel to, but slightly to the right of, the imaginary axis from $-\infty$ to $+\infty$. Due to the presence of the convergence factor $e^{0^{+}z}$, the integral along $C_1$ vanishes. Furthermore, in the zero-temperature limit ($\beta \to \infty$), the Bose distribution function vanishes: $n_B(z)=0$. Thus the integral along $C_2$ also equals zero.  }\label{fig3}  
\end{figure}

To regularize $\Omega_{\rm QF}$, we adopt a scheme similar to that used for pairing superfluids \cite{QF2D0}. Noting that $M_{11}^c(-q)=M_{22}^c(q)$, one has $\sum_{q}\ln \det M_{11}^c(q)=\sum_{q}\ln \det M_{22}^c(q)$. Then we perform analytic continuation of $M_{11}^c(\mathbf{q},\mathrm{i}v_n)$ to the entire complex plane, denoted as $M_{11}^c(\mathbf{q},z)$. If $\det M_{11}^c(\mathbf{q},z)\ne0$ in the left half plane ($\mathrm{Re} z<0 $), we can construct the integration contour shown in Fig. \ref{fig3} and prove 
\begin{align}
    &\frac{1}{2\pi \mathrm{i}}\oint_{C_1+C_2}n_B(z)\ln \det M_{11}^c(\mathbf{q},z)\mathrm{e}^{0^{+}z}dz\nonumber\\&=\frac{1}{\beta}\sum_{v_n}\ln \det M_{11}^c(\mathbf{q},\mathrm{i}v_n)=0,
\end{align}
where $n_B(z)=1/(\mathrm{e}^{\beta z}-1)$ is Bose distribution. 

Now we prove $\det M_{11}^c(\mathbf{q},z)\ne0$ in the left half of the complex plane. Since $M_{11}^c(\mathbf{q},z,\tilde{\mu}_Q)$ is a complex symmetric matrix, it suffices to show that its real part $\mathrm{Re}M_{11}^c(\mathbf{q},z,\tilde{\mu}_Q)$ is positive definite. Physically, a nonzero total momentum $\mathbf{q}$ corresponds to a moving quartet with kinetic energy. This indicates that $ \mathrm{Re}M_{11}^c(\mathbf{q},z,\tilde{\mu}_Q)$ becomes more positive definite than $\mathrm{Re}M_{11}^c(\mathbf{0},z,\tilde{\mu}_Q)$, leading to  $\det\mathrm{Re} M_{11}^c(\mathbf{q},z,\tilde{\mu}_Q)>\det \mathrm{Re}M_{11}^c(\mathbf{0},z,\tilde{\mu}_Q)$. Likewise, a larger $\tilde{\mu}_Q$, which corresponds to adding more quartets, should also give $\det\mathrm{Re} M_{11}^c(\mathbf{q},z,\tilde{\mu}_Q)>\det \mathrm{Re}M_{11}^c(\mathbf{q},z,0)$.  A rigorous analytic proof of these two monotonicities is difficult. However, in our system, we work in the deep binding limit where $\tilde{\mu}_Q\to0^+$, so we approximate $\det \mathrm{Re}M_{11}^c(\mathbf{q},z,\tilde{\mu}_Q)\approx\det \mathrm{Re}M_{11}^c(\mathbf{q},z,0)$. Then, for any nonzero vector $\mathbf{f}=(...,f_{\mathbf{k_2k_3}},...)$, we have $\mathbf{f}^T\mathrm{Re}M_{11}^c(\mathbf{q},z,0)\mathbf{f}\ge \mathbf{f}^T\mathrm{Re}M_{11}^c(\mathbf{q},0,0)\mathbf{f}$. The matrix $\mathrm{Re}M_{11}^c(\mathbf{q},0,0)$ corresponds to the few-body equation (\ref{qfewbodyequ}) with a macroscopic momentum, but with $E=E_4$ instead of $E_4+E_q$. Since the ground-state energy of a finite-momentum quartet satisfies $E(q)>E(0)=E_4$, and $M_{\rm FE}$ is positive semidefinite, the matrix $\mathrm{Re}M_{11}^c(\mathbf{q},0,0)$ is positive definite when $\mathbf{q}\ne\mathbf{0}$. Only at $\mathbf{q}=\mathbf{0}$ does the matrix become positive semidefinite.  However, for any nonzero vector $\mathbf{f}$, we have $\mathbf{f}^T\mathrm{Re}M_{11}^c(\mathbf{0},z,\tilde{\mu}_Q)\mathbf{f}>0$. Therefore, we conclude that the real part of $M_{11}^c(\mathbf{q},z,\tilde{\mu}_Q)$ is positive definite in the left half-plane, and consequently $\det M_{11}^c(\mathbf{q},z)\ne0$.

Based on the above discussion, we can rewrite $\Omega_{\mathrm{QF}}$ as
\begin{equation}
\Omega_{\mathrm{QF}}=\frac{1}{2\beta}\sum_{q}\ln\frac{\det M}{\det M_{\mathrm{DB}}^c}.
\end{equation}
At $T=0$, we replace the discrete Matsubara frequency sum with a continuous integral over an imaginary frequency, and then obtain 
\begin{equation}
\Omega_{\mathrm{QF}}=\frac{1}{2}\sum_{\mathbf{q}}\int_{-\infty}^{+\infty}\frac{\mathrm{d}\omega}{2\pi}\ln\frac{\det M}{\det M_{\mathrm{DB}}^c}.\label{OMEGAGF1}
\end{equation}

\section{Asymptotic behavior of $\Omega_{\mathrm{QF}}$}\label{sec:appendixasy}

Utilizing  $(M_{\mathrm{DB}}^{c})^{-1}=\mathrm{adj}(M_{\mathrm{DB}}^c)/\det M_{\mathrm{DB}}^c$,  $\ln (1+x)=-\sum_{n=1}^{\infty}(-x)^n/n$, and Eq.(\ref{detconvergent}) we can rewrite Eq.(\ref{OMEGAGF1}) as
\begin{align}
    \Omega_{\mathrm{QF}}=&-\frac{1}{2}\sum_{\mathbf{q}}\int_{-\infty}^{+\infty}\frac{\mathrm{d}\omega}{2\pi}\nonumber\\\times&\sum_{n=1}^{\infty} \frac{1}{n} 
    \{\frac{1}{2}\frac{\mathrm{tr}[\mathrm{adj}(M_{\mathrm{DB}}^{c})M_{\mathrm{ODB}}\mathrm{adj}(M_{\mathrm{DB}}^{c})M_{\mathrm{ODB}}]}{(\det M_{11}^c\det M_{22}^c)^2} \nonumber\\&-\frac{\mathrm{tr}[\mathrm{adj}(M_{\mathrm{DB}}^{c})M_{\mathrm{DB}}^r]}{\det M_{11}^c\det M_{22}^c}\}^n.
\end{align}
Now we analyze the asymptotic behavior of $\Omega_{\mathrm{QF}}$. We start from  the lowest-order term, i.e., $n=1$. 

For the first term, in the deep binding and low-energy limit, we can prove
\begin{align}
    &\frac{1}{2}\frac{\mathrm{tr}[\mathrm{adj}(M_{\mathrm{DB}}^{c})M_{\mathrm{ODB}}\mathrm{adj}(M_{\mathrm{DB}}^{c})M_{\mathrm{ODB}}]}{(\det M_{11}^c\det M_{22}^c)^2}\nonumber\\&\approx\frac{\tilde{\mu}_Q^2}{(q^2/2M_Q+\tilde{\mu}_Q)^2+\omega^2}.
\end{align}
The corresponding integral $I_{1r_1}$  gives
\begin{align}
    I_{1r_1}&=-\frac{1}{2}\sum_{\mathbf{q}}\int_{-\infty}^{+\infty}\frac{\mathrm{d}\omega}{2\pi} \frac{\tilde{\mu}_Q^2}{(q^2/2M_Q+\tilde{\mu}_Q)^2+\omega^2}\nonumber \\&= -\frac{\tilde{S}M_QE_{2b}}{8\pi^2m_r}\int_{0}^{+\infty} \mathrm{d}x\int_{0}^{+\infty}\mathrm{d}y\frac{z^2}{(x+z)^2+y^2},
\end{align}
where we have defined $x=E_q/E_{2b}$, $y=\omega/E_{2b}$ and $z=\tilde{\mu}_Q/E_{2b}$. Then we transform to polar coordinates $x=\rho \cos \theta$ and $y=\rho \sin \theta$ for convenience. Since we are primarily concerned with the contribution from the integral in the vicinity of the origin ($q=0$ and $\omega=0$), denoted as $I_{1r_1}^{\epsilon}$, we then have
\begin{align}
    I_{1r_1}^{\epsilon}&=-\frac{\tilde{S}M_QE_{2b}}{8\pi^2 m_r}\int_{0}^{\frac{\pi}{2} }\mathrm{d}\theta \int _{0}^{\epsilon }\rho \mathrm{d}\rho \frac{z^2}{\rho^2+2\rho z \cos \theta+z^2}\nonumber \\&\sim\frac{\tilde{S}M_QE_{2b}}{16\pi m_r} z^2\ln z.
\end{align}
Here we have utilized the following integral \cite{QF2D0}
\begin{equation}
    I_{mn}=\int_{0}^{\epsilon}\rho \mathrm{d}\rho \frac{\rho ^m}{(\rho ^2+2\zeta \rho \cos \varphi +\zeta^2)^n}.
\end{equation}
The properties of the integral $I_{mn}$ can be summarized as follows: For $m>2(n-1)$, the integral is finite; for $m=2(n-1)$, it diverges as $I_{mn}\sim -\ln \zeta$; and for $m<2(n-1)$,  $I_{mn} \sim 1/\zeta ^{2n-2-m}$ as $\zeta \to 0$.

For the second term, $M_{\mathrm{DB}}^r$ is a complex matrix. Therefore, we expand both its real and imaginary parts to the lowest order in $\tilde{\mu}_Q$, $q$ and $\omega$ in the deep binding and low-energy limit, i.e., $M_{11}^r\approx R+\mathrm{i}\omega I$ and $M_{22}^r\approx R-\mathrm{i}\omega I$. Then we get $\mathrm{tr}[\mathrm{adj}(M_{\mathrm{DB}}^{c})M_{\mathrm{DB}}^r]\approx 2\mathrm{tr}(\tilde{A}B)[(E_q+\tilde{\mu}_Q)\mathrm{tr}(\tilde{A}R)+\omega^2\mathrm{tr}(\tilde{A}I)]$, and the corresponding integral $I_{1r_2}$ reads
\begin{align}
    I_{1r_2}&=\sum_{\mathbf{q}}\int_{-\infty}^{+\infty}\frac{\mathrm{d}\omega}{2\pi} \frac{(E_q+\tilde{\mu}_Q)}{(E_q+\tilde{\mu}_Q)^2+\omega^2}\frac{\mathrm{tr}(\tilde{A}R)}{\mathrm{tr}(\tilde{A}B)}\nonumber\\&+\sum_{\mathbf{q}}\int_{-\infty}^{+\infty}\frac{\mathrm{d}\omega}{2\pi} \frac{\omega^2}{(E_q+\tilde{\mu}_Q)^2+\omega^2}\frac{\mathrm{tr}(\tilde{A}I)}{\mathrm{tr}(\tilde{A}B)}\label{I1r2}.
\end{align}
Because $M_{11}^r\approx R+\mathrm{i}\omega I \propto \tilde{\mu}_Q^2$ in the deep binding  limit, we have $\mathrm{tr}(\tilde{A}R)/\mathrm{tr}(\tilde{A}B)\propto \tilde{\mu}_Q^2$ and $\mathrm{tr}(\tilde{A}I)/\mathrm{tr}(\tilde{A}B)\propto \tilde{\mu}_Q^2$.  Then the first term on the right-hand side of Eq.(\ref{I1r2}) gives
\begin{align}
I_{1}^{\epsilon}&\propto z^2\int_{0}^{\epsilon}\rho\mathrm{d}\rho \frac{(\rho \cos \theta+z)}{\rho^2+2\rho z \cos \theta+z^2} \nonumber\\& \sim z^2(\mathrm{const}.+z\ln z),
\end{align}
while the second term in Eq.(\ref{I1r2}) yields the integral $I_2^{\epsilon}$:
\begin{equation}
    I_2^{\epsilon}\propto z^2\int_{0}^{\epsilon}\rho \mathrm{d}\rho \frac{\rho ^2 \sin ^2 \theta}{\rho ^2+2 \rho z \cos \theta +z^2} \propto z^2.
\end{equation}
Therefore, $I_{1r_2}^{\epsilon}$  contributes to the next-leading-order terms proportional to $\tilde{\mu}_Q^2$. 

Concerning the higher-order terms, due to the presence of $\tilde{\mu}_Q^{2n}$ factors, we only need to consider $I_{nr_1}$:
\begin{equation}
    I_{nr_1}\propto \sum_{\mathbf{q}}\int_{-\infty}^{+\infty}\mathrm{d}\omega  \frac{\tilde{\mu}_Q^{2n}}{[(q^2/2M_Q+\tilde{\mu}_Q)^2+\omega^2]^n},
\end{equation}
which contributes to the integral $I_{n}^{\epsilon}$
\begin{equation}
    I_n^{\epsilon}\propto\int _{0}^{\epsilon }\rho \mathrm{d}\rho \frac{z^{2n}}{(\rho^2+2\rho z \cos \theta+z^2)^n} \sim z^2.
\end{equation}

Based on the above discussions, we can conclude  that the asymptotic behavior of $\Omega_{\mathrm{QF}}/E_{2b}$ is dominated by $\tilde{\mu}_Q^2 \ln (\tilde{\mu}_Q/E_{2b})/E_{2b}^2$, with coefficient given by $\tilde{S}M_Q/16\pi m_r $.

\section{Numerical simulation of $\Omega_{\mathrm{QF}}$}\label{sec:appendixnum_cal}

To numerically evaluate $\Omega_{\mathrm{QF}}$, we need to compute the functional determinant $\det M$ and $\det M_{\mathrm{DB}}^c$. Because $M$ contains off-diagonal elements, we discretize the momenta $\mathbf{k}$ on a uniform Cartesian grid. This discretization introduces three parameters: the grid size $n_x$, $n_y$ and the momentum cutoff $k_c$. Due to computational limitations, we fix $n_x=n_y=10$ and impose the boundary condition that $\Delta_{\mathbf{k_2k_3}}=0$ whenever any of the momenta satisfies $|k_{i,x}|>k_c$ or $|k_{i,y}|>k_c$. To reproduce the known analytical results, we choose the optimal $k_c$ such that the second derivative of $\det M_{\mathrm{FE}}(q)$ with respect to $q$ at $q\rightarrow 0$ equals $\mathrm{tr}(\tilde{A}B)/2M_Q$, as proved in appendix C.

Following this procedure, we obtain $k_ca=3.3$ for Li-K and $k_ca=3.6$ for Li-Cr. The gap equation (\ref{gapequation_T0}) is then solved on the Cartesian grid for various $\tilde{\mu}_Q$. For a given nonzero $\mathbf{q}$, we obtain $\Delta_{\mathbf{k_2+q,k_3}}$ via a two-dimensional linear interpolation. Since $\Delta_{\mathbf{k}_2\mathbf{k_3}}$ is stored as an antisymmetric matrix, row interpolation (which gives $\Delta_{\mathbf{k}_3,\mathbf{k}_2+\mathbf{q}}$) and column interpolation (which gives $\Delta_{\mathbf{k}_2+\mathbf{q},\mathbf{k}_3}$) automatically yield results related by antisymmetry. We choose row interpolation in our numerical evaluation.

Next, to handle the determinant of the complex matrix $M=R+\mathrm{i}I$, we employ its real representation $\bar{M} = \begin{pmatrix} R & -I \\ I & R \end{pmatrix}$, which satisfies $\det \bar{M}=|\det M|^2$. Using this relation, we rewrite the fluctuation thermodynamic potential as 
\begin{equation}
     \Omega_{\mathrm{QF}}=\frac{1}{2}\sum_{\mathbf{q}}\int_{0}^{+\infty}\frac{\mathrm{d}\omega}{2\pi}\ln\frac{\det \bar{M}}{\det \bar{M}_{\mathrm{DB}}^c}.
 \end{equation}
We finally use Gauss-Laguerre quadrature to numerically perform the integrations over $\mathbf{q}$ and $\omega$, which is well suited for integrals exhibiting the expected asymptotic behavior $\tilde{\mu}_Q^2\ln (\tilde{\mu}_Q/E_{2b})$. The results obtained with this numerical scheme are presented in the main text (see Fig.\ref{fig_num}).


\begin{thebibliography}{99}

\bibitem{review_1}I. Bloch, J. Dalibard, and W. Zwerger, \textit{Many-body physics with ultracold gases}, Rev. Mod. Phys. \textbf{80}, 885 (2008).
\bibitem{review_2}S. Giorgini, L. P. Pitaevskii,  and S. Stringari, \textit{Theory of ultracold atomic Fermi gases}, Rev. Mod. Phys. \textbf{80}, 1215 (2008).
\bibitem{review_nuclear1}D. J. Dean and M. Hjorth-Jensen, \textit{Pairing in nuclear systems: from neutron stars to finite nuclei}, Rev. Mod. Phys. \textbf{75}, 607 (2003).
\bibitem{review_nuclear2}M. G. Alford, A. Schmitt, K. Rajagopal, and T. Sch{\" a}fer, \textit{Color superconductivity in dense quark matter}, Rev. Mod. Phys. \textbf{80}, 1455 (2008).

\bibitem{3D1}M. Greiner, C. A. Regal, and D. S. Jin, \textit{Emergence of a molecular Bose–Einstein condensate from a Fermi gas}, Nature \textbf{426}, 537 (2003).
\bibitem{3D2}M. W. Zwierlein, C. A. Stan, C. H. Schunck, S. M. F. Raupach, S. Gupta, Z. Hadzibabic, and W. Ketterle, \textit{Observation of Bose-Einstein condensation of molecules}, Phys. Rev. Lett. \textbf{91}, 250401 (2003).
\bibitem{3D3}S. Jochim, M. Bartenstein, A. Altmeyer, G. Hendl, S. Riedl, C. Chin, J. H. Denschlag, and R. Grimm, \textit{Bose-Einstein condensation of molecules}, Science \textbf{302}, 2101 (2003).
\bibitem{3D4}M. W. Zwierlein, J. R. Abo-Shaeer, A. Schirotzek, C. H. Schunck, and W. Ketterle, \textit{Vortices and superfluidity in a strongly interacting Fermi gas}, Nature \textbf{435}, 1047 (2005).
\bibitem{3D5}J. Joseph, B. Clancy, L. Luo, J. Kinast, A. Turlapov, and J. E. Thomas, \textit{Measurement of sound velocity in a Fermi gas near a Feshbach Resonance}, Phys. Rev. Lett. \textbf{98}, 170401 (2007).
\bibitem{3D6}D. E. Miller, J. K. Chin, C. A. Stan, Y. Liu, W. Setiawan, C. Sanner, and W. Ketterle, \textit{Critical velocity for superfluid flow across the BEC-BCS crossover}, Phys. Rev. Lett. \textbf{99}, 070402 (2007).
\bibitem{3D7}C. H. Schunck, Y. Shin, A. Schirotzek, and W. Ketterle, \textit{Determination of the fermion pair size in a resonantly interacting superfluid}, Nature \textbf{454}, 739 (2008).
\bibitem{3D8}N. Navon, S. Nascimb\`ene, F. Chevy, and C. Salomon, \textit{The equation of state of a low-temperature Fermi gas with tunable interactions}, Science \textbf{328}, 729 (2010).
\bibitem{3D9}A. Perali, F. Palestini, P. Pieri, G. C. Strinati, J. T. Stewart, J. P. Gaebler, T. E. Drake, and D. S. Jin, \textit{Evolution of the normal state of a strongly interacting Fermi gas from a pseudogap to a molecular Bose gas}, Phys. Rev. Lett. \textbf{106}, 060402 (2011).
\bibitem{3D10}C. Cao, E. Elliott, J. Joseph, H. Wu, J.Petricka, T. Sch\"afer, and J. E. Thomas, \textit{Universal quantum viscosity in a unitary Fermi gas}, Science \textbf{331}, 58 (2011).
\bibitem{3D11}M. J. H. Ku, A. T. Sommer, L. W. Cheunk, and M. W. Zwierlein, \textit{Revealing the superfluid lambda transition in the universal thermodynamics of a unitary Fermi gas}, Science \textbf{335}, 563 (2012).
\bibitem{3D12}L. A. Sidorenkov, M. K. Tey, R. Grimm, Y.-H. Hou, L. Pitaevskii, and S. Stringari, \textit{Second sound and the superfluid fraction in a Fermi gas with resonant interactions}, Nature \textbf{498}, 78 (2013).
\bibitem{3D13}W. Weimer, K. Morgener, V. P. Singh, J. Siegl, K. Hueck, N. Luick, L. Mathey and H. Moritz, \textit{Critical velocity in the BEC-BCS crossover}, Phys. Rev. Lett. \textbf{114}, 095301 (2015).
\bibitem{3D14}S. Hoinka, P. Dyke, M. G. Lingham, J. J. Kinnumen, G. M. Bruun, and C. J. Vale, \textit{Goldstone mode and pair-breaking excitations in atomic Fermi superfluids}, Nat. Phys. \textbf{13}, 942 (2017).
\bibitem{3D15}H. Biss, L. Sobirey, N. Luick, M. Bohlen, J. J. Kinnunen, G. M. Bruun, T. Lompe, and H. Moritz, \textit{Excitation spectrum and superfluid gap of an ultracold Fermi gas}, Phys. Rev. Lett. \textbf{128}, 100401 (2022).
\bibitem{3D16}X. Li, S. Wang, X. Luo, Y.-Y. Zhou, K. Xie, H.-C. Shen, Y.-Z. Nie, Q. Chen, H. Hu, Y.-A. Chen, X.-C. Yao, and J.-W. Pan, \textit{Observation and quantification of the pseudogap in unitary Fermi gases}, Nature \textbf{626}, 288 (2024).

\bibitem{2D1}K. Martiyanov, V. Makhalov, and A. Turlapov, \textit{Obseration of a two-dimensional Fermi gas of atoms}, Phys. Rev. Lett. \textbf{105}, 030404 (2010).
\bibitem{2D2}B. Fr\"ohlich, M. Feld, E. Vogt, M. Koschorreck, W. Zwerger, and M. K\"ohl, \textit{Radio-Frequency spectroscopy of a strongly interacting two-dimensional Fermi gas}, Phys. Rev. Lett. \textbf{106}, 105301 (2011).
\bibitem{2D3}P. Dyke, E. D. Kuhnle, S. Whitlock, H. Hu, M. Mark, S. Hoinka, M. Lingham, P. Hannaford, and C. J. Vale, \textit{Crossover from 2D to 3D in a weakly interacting Fermi gas}, Phys. Rev. Lett. \textbf{106}, 105304 (2011).
\bibitem{2D4}M. Feld, B. Fr\"ohlich, E. Vogt, M. Koschorreck, and M. K\"ohl, \textit{Observation of a pairing pseudogap in a two-dimensional Fermi gas}, Nature \textbf{480}, 75 (2011).
\bibitem{2D5}V. Makhalov, K. Martiyanov, and A. Turlapov, \textit{Ground-state pressure of quasi-2D Fermi and Bose gases}, Phys. Rev. Lett. \textbf{112}, 045301 (2014).
\bibitem{2D6}M. G. Ries, A. N. Wenz, G. Z\"urn, L. Bayha, I. Boettcher, D. Kedar, P. A. Murthy, M. Neidig, T. Lompe, and S. Jochim, \textit{Observation of pair condensation in the quasi-2D BEC-BCS crossover}, Phys. Rev. Lett. \textbf{114}, 230401 (2015).
\bibitem{2D7}P. A. Murthy, I. Boettcher, L. Bayha, M. Holzmann, D. Kedar, M. Neidig, M. G. Ries, A. N. Wenz, G. Z\"urn, and S. Jochim, \textit{Observation of the Berezinskii-Kosterlitz-Thouless phase transition in an ultracold Fermi gas}, Phys. Rev. Lett. \textbf{115}, 010401 (2015).
\bibitem{2D8}I. Boettcher, L. Bayha, D. Kedar, P. A. Murthy, M. Neidig, M. G. Ries, A. N. Wenz, G. Z\"urn, S. Jochim, and T. Enss, \textit{Equation of state of ultracold fermions in the 2D BEC-BCS crossover region}, Phys. Rev. Lett. \textbf{116}, 045303 (2016).
\bibitem{2D9}P. A. Murthy, M. Neidig, R. Klemt, L. Bayha, I. Boettcher, T. Enss, M. Holten, G. Z\"urn, P. M. Preiss, and S. Jochim, \textit{High-temperature pairing in a strongly interacting two-dimensional Fermi gas}, Science \textbf{359}, 452 (2018).
\bibitem{2D10}L. Sobirey, N. Luick, M. Bohlen, H. Biss, H. Moritz, and T. Lompe, \textit{Observation of superfluidity in a strongly correlated two-dimensional Fermi gas}, Science \textbf{372}, 844 (2021).
\bibitem{2D11} M. Holten, L. Bayha, K. Subramanian, S. Brandstetter, C. Heintze, P.. Lunt, P. M. Preiss, and S. Jochim, \textit{Observation of Cooper pairs in a mesoscopic two-dimensional Fermi gas}, Nature \textbf{606}, 287 (2022).

\bibitem{MC3D1}G. E. Astrakharchik, J. Boronat, J. Casulleras, and S. Giorgini, \textit{Equation of state of a Fermi gas in the BEC-BCS crossover: a quantum Monte Carlo study}, Phys. Rev. Lett. \textbf{93}, 200404 (2004).
\bibitem{MC3D2}M. M. Forbes, S. Gandolfi, and A. Gezerlis, \textit{Resonantly interacting fermions in a box}, Phys. Rev. Lett. \textbf{106}, 235303 (2011).
\bibitem{MC3D3}S. Gandolfi, K. E. Schmidt, and J. Carlson, \textit{BEC-BCS crossover and universal relations in unitary Fermi gases}, Phys. Rev. A \textbf{83}, 041601(R) (2011).
\bibitem{MC3DT1}O. Goulko, and M. Wingate, \textit{Numerical study of the unitary Fermi gas across the superfluid transition}, Phys. Rev. A \textbf{93}, 053604 (2016).
\bibitem{MC3DT2}S. Jensen, C. N. Gilbreth, and Y. Alhassid, \textit{Pairing correlations across the superfluid phase transition in the unitary Fermi gas}, Phys. Rev. Lett. \textbf{124}, 090604 (2020).
\bibitem{MC3DT3}A. Richie-Halford, J. E. Drut, and A. Bulgac, \textit{Emergence of a pseudogap in the BCS-BEC crossover}, Phys. Rev. Lett. \textbf{125}, 060403 (2020).
\bibitem{MC2D1}G. Bertaina, and S. Giorgini, \textit{BCS-BEC crossover in a two-dimensional Fermi gas}, Phys. Rev. Lett. \textbf{106}, 110403 (2011).
\bibitem{MC2D2}H. Shi, S. Chiesa, and S. Zhang, \textit{Ground-state properties of strongly interacting Fermi gases in two dimensions}, Phys. Rev. A \textbf{92}, 033603 (2015).
\bibitem{MC2D3}A. Galea, H. Dawkins, S. Gandolfi, and A. Gezerlis, \textit{Diffusion Monte Carlo study of strongly interacting two-dimensional Fermi gases}, Phys. Rev. A \textbf{93}, 023602 (2016).
\bibitem{MC2DT1}E. R. Anderson, and J. E. Drut, \textit{Pressure, compressibility, and contact of the two-dimensional attractive Fermi gas}, Phys. Rev. Lett. \textbf{115}, 115301 (2015).
\bibitem{MC2DT2}Y.-Y. He, H. Shi, and S. Zhang, \textit{Precision many-body study of the Berezinskii-Kosterlitz-Thouless transition and temperature-dependent properties in the two-dimensional Fermi gas}, Phys. Rev. Lett. \textbf{129}, 076403 (2022).
\bibitem{MC2DT3}S. Ramachandran, S. Jensen, and Y. Alhassid, \textit{Pseudogap effects in the strongly correlated regime of the two-dimensional Fermi gas}, Phys. Rev. Lett. \textbf{133}, 143405 (2024).
\bibitem{MCmassimba1}A. Gezerlis, S. Gandolfi, K. E. Schmidt, and J. Carlson, \textit{Heavy-light fermion mixtures at unitarity}, Phys. Rev. Lett. \textbf{103}, 060403 (2009).
\bibitem{MCmassimba2}J. Braun, J. E. Drut, and D. Roscher, \textit{Zero-temperature equation of state of mass-imbalanced resonant fermi gases}, Phys. Rev. Lett. \textbf{114}, 050404 (2015).
\bibitem{MCmassimba3}S. Gandolfi, R. Curry, and A. Gezerlis, \textit{BCS-BEC crossover of the strongly interacting ${}^6\textrm{Li}-{}^{40}\textrm{K}$ mixture}, Phys. Rev. A \textbf{110}, 043320 (2024).

\bibitem{NSR}P. Nozi\`eres, and S. Schmitt-Rink, \textit{Bose condensation in an attractive fermion gas: from weak to strong coupling superconductivity}, J. Low Temp. Phys. \textbf{59}, 195 (1985).
\bibitem{Melo1}C. A. R. S\'a de Melo, M. Randeria, and J. R. Engelbrecht, \textit{Crossover from BCS to Bose superconductivity: transition temperature and time-dependent Ginzburg-Landau theory}, Phys. Rev. Lett. \textbf{71}, 3202 (1993).
\bibitem{Haussmann1}R. Haussmann, \textit{Properties of a Fermi liquid at the superfluid transition in the crossover region between BCS superconductivity and Bose-Einstein condensation}, Phys. Rev. B \textbf{49}, 12975 (1994).
\bibitem{Ohashi1}Y. Ohashi, and A. Griffin, \textit{Superfluidity and collective modes in a uniform gas of Fermi atoms with a Feshbach resonance}, Phys. Rev. A \textbf{67}, 063612 (2003).
\bibitem{Strinati1}P. Pieri, L. Pisani, and G. C. Strinati, \textit{BCS-BEC crossover at finite temperature in the broken-symmetry phase}, Phys. Rev. B \textbf{70}, 094508 (2004).
\bibitem{ChenQijin}Q. Chen, J. Stajic, S. Tan, and K. Levin, \textit{BCS-BEC crossover: from high temperature superconductors to ultracold superfluids}, Phys. Rep. \textbf{412}, 1 (2005).
\bibitem{Ohashi2}E. Taylor, A. Griffin, N. Fukushima, and Y. Ohashi, \textit{Paring fluctuation and the superfluid density through the BCS-BEC crossover}, Phys. Rev. A \textbf{74}, 063626 (2006).
\bibitem{Haussmann2}R. Haussmann, W. Rantner, S. Cerrito, and W. Zwerger, \textit{Thermodynamics of the BCS-BEC crossover}, Phys. Rev. A \textbf{75}, 023610 (2007).
\bibitem{Ohashi3}R. Watanabe, S. Tsuchiya, and Y. Ohashi, \textit{Superfluid density of states and pseudogap phenomenon in the BCS-BEC crossover regime of a superfluid Fermi gas}, Phys. Rev. A \textbf{82}, 043630 (2010).
\bibitem{Ohashi4}H. Tajima, P. van Wyk, R. Hanai, D. Kagamihara, D. Inotani, M. Horikoshi, and Y. Ohashi, \textit{Strong-coupling corrections to ground-state properties of a superfluid Fermi gas}, Phys. Rev. A \textbf{95}, 043625 (2017).
\bibitem{Strinati2}M. Pini, P. Pieri, and G. C. Strinati, \textit{Fermi gas throughout the BCS-BEC crossover: Comparative study of t-matrix approaches with various degrees of self-consistency}, Phys. Rev. B \textbf{99}, 094502 (2019).

\bibitem{Melo2}J. R. Engelbrecht, M. Randeria, and C. A. R. S\'a de Melo, \textit{BCS to Bose crossover: broken-symmetry state}, Phys. Rev. B \textbf{55}, 15153 (1997).
\bibitem{HuHui1}H. Hu, X.-J. Liu and P. D. Drummond, \textit{Equation of state of a superfluid Fermi gas in the BCS-BEC crossover}, Europhys. Lett. \textbf{74}, 574 (2006).
\bibitem{Randeria}R. B. Diener, R. Sensarma, and M. Randeria, \textit{Quantum fluctuations in the superfluid state of the BCS-BEC crossover}, Phys. Rev. A \textbf{77}, 023626 (2008).
\bibitem{QF2D0}L. He, H. L\"u, G. Cao, H. Hu and X.-J. Liu, \textit{Quantum fluctuations in the BCS-BEC crossover of two-dimensional Fermi gases}, Phys. Rev. A \textbf{92}, 023620 (2015).
\bibitem{QF2DBKT}B. C. Mulkerin, L. He, P. Dyke, C. J. Vale, X.-J. Liu, and H. Hu, \textit{Superfluid density and critical velocity near the Berezinskii-Kosterlitz-Thouless transition in a two-dimensional strongly interacting Fermi gas}, Phys. Rev. A \textbf{96}, 053608 (2017).
\bibitem{Salasnich}G. Bighin, and L. Salasnich, \textit{Finite-temperature quantum fluctuations in two-dimensional Fermi superfluids}, Phys. Rev. B \textbf{93}, 014519 (2016).

\bibitem{Wu1}C. Wu, \textit{Competing orders in one-dimensional spin-3/2 fermionic systems}, Phys. Rev. Lett. \textbf{95}, 266404 (2005).
\bibitem{quartet_nuclear1}G. R\"opke, A. Schnell, P. Schuck, and P. Nozi\`eres, \textit{Four-particle condensate in strongly coupled fermion systems}, Phys. Rev. Lett. \textbf{80}, 3177 (1998).
\bibitem{quartet_nuclear2}R. A. Sen'kov, and V. G. Zelevinsky, \textit{Unified BCS-like model of pairing and alpha-correlations}, Phys. At. Nucl. \textbf{74}, 1267 (2011).
\bibitem{quartet_nuclear3}V. V. Baran, and D. S. Delion, \textit{A quartet BCS-like theory}, Phys. Lett. B \textbf{805}, 135462 (2020).
\bibitem{quartet_nuclear4}V. V. Baran, D. R. Nichita, D. Negrea, D. S. Delion, N. Sandulescu, and P. Schuck, \textit{Bridging the quartet and pair pictures of isovector proton-neutron pairing}, Phys. Rev. C \textbf{102}, 061301(R) (2020).
\bibitem{Tajima1}Y. Guo, H. Tajima, and H. Liang, \textit{Cooper quartet correlations in infinite symmetric nuclear matter}, Phys. Rev. C \textbf{105}, 024317 (2022).
\bibitem{Tajima2}Y. Guo, H. Tajima, and H. Liang, \textit{Biexciton-like quartet condensates in an electron-hole liquid}, Phys. Rev. Research \textbf{4}, 023152 (2022).
\bibitem{Babaev1}E. Babaev, A. Sudb$\o$, and N. W. Ashcroft, \textit{A superconductor to superfluid phase transition in liquid metallic hydrogen}, Nature \textbf{431}, 666 (2004).
\bibitem{Kivelson}E. Berg, E. Fradkin, and S. A. Kivelson, \textit{Charge-4$e$ superconductivity from pair-density-wave order in certain high-temperature superconductors}, Nat. Phys. \textbf{5}, 830 (2009).
\bibitem{Babaev2}E. V. Herland, E. Babaev, and A. Sudb$\o$, \textit{Phase transitions in a three dimensional $U(1)\times U(1)$ lattice London superconductor: Metallic superfluid and charge-4$e$ superconducting states}, Phys. Rev. B \textbf{82}, 134511 (2010).
\bibitem{Yao1}Y.-F. Jiang, Z.-X. Li, S. A. Kivelson, and H. Yao, \textit{Charge-4$e$ superconductors: A Majorana quantum Monte Carlo study}, Phys. Rev. B \textbf{95}, 241103(R) (2017).
\bibitem{Fu}R. M. Fernandes, and L. Fu, \textit{Charge-4$e$ superconductivity from multicomponent nematic pairing: Application to twisted bilayer graphene}, Phys. Rev. Lett. \textbf{127}, 047001 (2021).
\bibitem{Yao2} S.-K. Jian, Y. Huang, and H. Yao, \textit{Charge-4$e$ superconductivity from nematic superconductors in two and three dimensions}, Phys. Rev. Lett. \textbf{127}, 227001 (2021).
\bibitem{Babaev3}V. Grinenko \textit{et al.}, \textit{State with spontaneously broken time-reversal symmetry above the superconducting phase transition}, Nat. Phys. \textbf{17}, 1254 (2021).
\bibitem{charge4e_expt}J. Ge, P. Wang, Y. Xing, Q. Yin, A. Wang, J. Shen, H. Lei, Z. Wang, and J. Wang, \textit{Charge-4$e$ and charge-6e flux quantization and higher charge superconductivity in kagome superconductor ring devices}, Phys. Rev. X \textbf{14}, 021025 (2024).
\bibitem{HuJiangping}P. Li, K. Jiang, and J. Hu, \textit{Charge 4$e$ superconductor: A wavefunction approach}, Science Bulletin \textbf{69}, 2328 (2024).
\bibitem{WangYuxuan}Y.-M. Wu, and Y. Wang, \textit{d-wave charge-4$e$ superconductivity from fluctuating pair density waves}, npj Quantum Materials \textbf{9}, 66 (2024).
\bibitem{Wu2}Z.-Q. Gao, Y.-Q. Wang, H. Yang, and C. Wu, \textit{Topological charge-2$ne$ superconductors}, arXiv:2512.21325 [cond-mat.str-el].
\bibitem{Babaev4}A. Samoilenka, and E. Babaev, \textit{Microscopic theory of electron quadrupling condensates}, Phys. Rev. Research \textbf{8}, 013139 (2026).
\bibitem{Jian}Z.-Q. Wan, H. Jiang, X. Zou, S. Zhang, and S.-K. Jian, \textit{Quantum charge-4$e$ superconductivity and deconfined pseudocriticality in the attractive SU(4) Hubbard model}, 	arXiv:2604.14289 [cond-mat.str-el].


\bibitem{Cui2} R. Liu, W. Wang, and X. Cui, \textit{Quartet superfluid in two-dimensional mass-imbalanced Fermi mixtures}, Phys. Rev. Lett. \textbf{131}, 193401 (2023).
\bibitem{Blume}D. Blume, \textit{Universal Four-Body States in Heavy-Light Mixtures with a Positive Scattering Length}, Phys. Rev. Lett. \textbf{109}, 230404 (2012). 
\bibitem{Petrov} B. Bazak and D. S. Petrov, \textit{Five-Body Efimov Effect and Universal Pentamer in Fermionic Mixtures}, Phys. Rev. Lett. \textbf{118}, 083002 (2017).
\bibitem{Parish}J. Levinsen and M. M. Parish, \textit{Bound States in a Quasi-Two-Dimensional Fermi Gas}, Phys. Rev. Lett. \textbf{110}, 055304 (2013).
\bibitem{Cui1}R. Liu, C. Peng, and X. Cui, \textit{Universal tetramer and pentamer bound states in two-dimensional fermionic mixtures}, Phys. Rev. Lett. \textbf{129}, 073401 (2022).




\bibitem{2DBoson}C. Mora, and Y. Castin, \textit{Ground state energy of the two-dimensional weakly interacting Bose gas: First correction beyond Bogoliubov theory}, Phys. Rev. Lett. \textbf{102}, 180404 (2009).

\bibitem{BKTestimate} N. Prokof'ev, O. Ruebenacker, and B. Svistunov, \textit{Critical point of a weakly interacting two-dimensional Bose gas}, Phys. Rev. Lett. \textbf{87}, 270402 (2001).

\bibitem{Data}W. Wang, Y. Wang, and X. Cui, \textit{Data for quantum fluctuations in a quartet superfluid of two-dimensional Fermi mixtures}, https://doi.org/10.6084/m9.figshare.32287221.v3.
	
\end{thebibliography}
\end{document}